\documentclass[12pt]{article}
\usepackage{epsfig, amssymb}
\usepackage{graphicx,epsfig} 
\begin{document}
\begin{titlepage}
\thispagestyle{empty}
\begin{flushright}
UK/10-13
\end{flushright}

\bigskip

\begin{center}
\noindent{\Large \textbf
{Small Amplitude Forced Fluid Dynamics from Gravity at $T=0$}}\\
\vspace{2cm} \noindent{
Jae-Hyuk Oh\footnote{e-mail:jaehyukoh@uky.edu}}

\vspace{1cm}
  {\it
 Department of Physics and Astronomy, \\
 University of Kentucky, Lexington, KY 40506, USA\\
 }
\end{center}

\vspace{0.3cm}
\begin{abstract}
The usual derivative expansion of gravity duals of charged fluid dynamics is known to break down in the zero temperature limit. In this case, the fluid-gravity duality is not understood precisely. We explore this problem for a zero temperature charged fluid driven by a low frequency, small amplitude and spatially homogeneous external force. In the gravity dual, this corresponds to time dependent boundary value of the dilaton. We calculate the bulk solution for the dilaton and the leading backreaction to the metric and the gauge fields using the modified low frequency expansion of \cite{Hong1}. The resulting solutions are regular everywhere, establishing fluid-gravity duality to this order.

\end{abstract}
\end{titlepage}
\newpage

\tableofcontents
\newpage
\section{Introduction}

The AdS/CFT correspondence has provided useful insight into the
dynamics of strongly coupled quantum field theories, particularly
nonabelian gauge theories. Recently this has led to a fluid-gravity
correspondence which provides a study of conformal fluid dynamics, an
effective description of strongly coupled conformal field theory at
long wavelengths in local equilibrium. In fact there is a precise
mathematical connection between a long distance limit of the Einstein
gravity and its holographic dual fluid dynamics
\cite{minwalla1,minwalla2,Hansen1,Nabamita1,Son1,Baier1}. Explicitly, it has been
demonstrated that certain 
deformations of asymptotically $AdS_{d+1}$ black branes
which are slowly varying along the boundary directions (but can have
$O(1)$ amplitudes) 
provide dual descriptions of solutions of the equations of fluid
dynamics.

In fluid dynamics, there are local thermodynamic quantities such
as local temperature, chemical potentials, $R$-charges or $d$-velocity,
which are slowly varying along the boundary directions 
compared to effective equilibrium length scale of the fluid, the mean free
path $l_{mfp}$. 
Bulk dual solutions of this inhomogeneous fluids are
constructed order by order in derivatives respect to boundary
coordinate.

More precisely, the asymptotically $AdS$ space is foliated into a
collection of tubes, each characterized by a value of the boundary
coordinate.  Each tube is centered about a radial ingoing null
geodesic starting from $AdS$ boundary. The width of each tube in the
boundary direction is smaller than the scale of dual fluid
dynamics. The gravity solution is developed locally in each tube,
which turns out to be black brane with local thermodynamic quantities
of the boundary fluid. The global geometry is constructed by gluing the
tubes, in which the local thermodynamic quantities change along the
boundary direction.

One crucial aspect of this construction is ultra-locality. The local
thermodynamic quantities and metric corrections are expanded around a
point on the boundary, which may be chosen to be $x^{\mu}$=0. For some
local thermodynamic quantity $q_i (x^\mu)$,
\begin{equation}
q_{i}(x^{\mu})=q_{i}(0)+x^{\mu}\partial_{\mu}q_{i}(0)+
\frac{1}{2}x^{\mu}x^{\nu}\partial_{\mu}\partial_{\nu}q_{i}(0)+...
\end{equation}
In equilibrium these quantities are independent of $x^\mu$ so that all
the higher terms vanish. The bulk then corresponds to some static
black brane.
These quantities appear in the bulk metric and other fields,
generically denoted by $g(r,x^\mu)$, where $r$ is the AdS radial
coordinate. One then expands
\begin{equation}
g(r,x^{\mu})=g^{(0)}(r,x^{\mu})+g^{(1)}(r,x^{\mu})+g^{(2)}(r,x^{\mu})+...,
\end{equation}
where $g^{(0)}(r,x^{\mu})$ is the bulk field obtained by replacing
$q_i$ in the equilibrium solution by
$q_i(x^\mu)$. $g^{(0)}(r,x^{\mu})$ is clearly not a solution of the
bulk equations of motion. The higher order terms $g^{(n)}(r,x^\mu)$
are then determined by requiring that the full $g (r,x^\mu)$ solves
the bulk equations of motion. The expansion above then constitutes a
derivative expansion of a bulk solution. The equation satisfied by
$g^{(n)}$ is of the form
\begin{equation}
\label{ultra-local operator}
H\left( g^{(0)}(q_{i}(0)) \right)g^{(n)}(r,x^{\mu})=s_{n}.
\end{equation}
$H$ is a linear differential operator in the second order in the
radial variable.  $s_{n}$ is a source term from $m$th order corrections
where $m < n$. The operator does not contain any derivative in the
boundary directions, since this would produce terms of higher order.
In other words, in this derivative expansion, $H$ becomes ultra-local
operator. The corrections $g^{(n)}(r,x^\mu)$ are required to be
regular everywhere outside the black brane horizon. They are also
required to fall of sufficiently fast at the boundary $r = \infty$, so
that they do not lead to additional sources in the boundary theory.

In \cite{Hansen1,Nabamita1}, this program has been carried out for
conformal fluids carrying a global R charge. However, it was found
that in the tubes where the local temperature is zero, i.e. the black
hole locally looks like an extremal black hole, the solutions are
singular at the horizon.  These singularities seem to be genuine in
that those possibly cause singularities in the curvature invariants at
black brane horizon. 

The failure of the derivative expansion method of deriving the bulk
solution in these tubes does not agree with expectations from the dual
fluid dynamics. In the dual fluid dynamics, the effective equilibrium
length scale is the mean free path, which is given by $l_{mfp} \sim
\frac{\eta}{\epsilon}$ \cite{minwalla3,Son2}. $\eta$ is shear
viscosity and $\epsilon$ is energy density. For the fluids described
by the gravity solutions in \cite{minwalla1,minwalla2,Hansen1,Nabamita1,Son1,Son3,Makoto1},
$\eta=\frac{s}{4\pi}$, where $s$ is entropy density. Consequently,
$l_{mfp} \sim \frac{s}{4\pi \epsilon}$. In a tube where the local
geometry becomes extremal, the zeroth order gravity solutions in the
derivative expansion in these papers have finite entropy and energy
densities.  This implies that there should be a reasonable fluid
dynamics even at zero temperature.

These divergences appear even in the linearized regime, i.e small
amplitude {\em and} small frequency perturbations around an extremal
black brane. Consider for example a linearized scalar field in the
extremal background. It has been shown in \cite{Hong1} that in this
case performing a small frequency expansion in the scalar field
equation is not straight-forward. Technically, this is because
the equation contains terms which have powers of the frequency
multiplied by functions which blow up at the horizon. This may be
traced to the fact that the components of the background metric have
double zeroes or poles at the horizon as opposed to single zeroes or
poles for a geometry at finite temperature.

A clue to resolve this problem is obtained by observing that the near horizon
geometry of the charged black brane becomes $AdS_{2}\times
R^{d-1}$, where $d+1$ becomes bulk space-time dimension. For the near
horizon region, it is natural to introduce a new radial coordinate
$\zeta \sim \frac{\omega}{r-r_{0}}$, where $\omega$ is frequency of
the scalar field, $r$ is radial coordinate of the black brane, and
$r_{0}$ is the horizon of the black brane. Motivated by the fact 
that in the near horizon region, the equations involve $\zeta$ 
with no further $\omega$ dependence,  the small frequency expansion 
is then obtained by writing everything in terms of $\zeta$ and then
expanding in powers of $\omega$. To lowest order in this
modified small frequency expansion, the equations are then solved
in two regions : (i) the inner region close to the horizon, which is 
defined as $\zeta \rightarrow \infty$ with
$\frac{\omega}{\zeta} \rightarrow 0$,
and (ii) the outer region which corresponds
to small $\zeta$ or $r \gg r_0$. It is essential to use the coordinate $\zeta$
in the inner region rather than $r$. The solution in the full space is
then obtained by matching the inner and outer region solutions.
By construction, a solution which is regular at the horizon exists.
This technique has been used to obtained various response
functions by solving the linearized gravity equations and gauge 
field equations \cite{Edalati1,Edalati2}.

As will be clarified below, this modified low frequency expansion
implies that the tubewise approximation used in obtaining gravity
duals of boundary fluid flows breaks down. This is because the change
of radial coordinates from $r$ to $\zeta$ involves the frequency, so
that in position space this implies a non-local (on the boundary)
redefinition of fields. This reorganizes the low frequency expansion in which differential operators in the equations are not ultra-local any longer. The implications of this fact for fluid
mechanics are not clear at the moment. 

In this paper we consider a related problem where similar divergences
appear and show how these get resolved. We consider four dimensional
Einstein-Maxwell-dilaton theory with a negative cosmological
constant. All bulk fields are spatially homogeneous but time
dependent. The dilaton has a nonzero boundary value and the velocity
of the boundary fluid is zero. First we consider slowly varying
deformations of a charged black hole in this geometry in the presence of
the dilaton source. These deformations are {\em entirely} due to the
dilaton source.  The setup is similar to that of \cite{Awad:2009bh}
where a time dependent boundary value of the dilaton evolves an
initially pure AdS geometry into that with a space-like region of
large curvature. In the sense of fluid dynamics, our set up is a natural extension of \cite{minwalla2} to $R$-charged fluid with vanishing velocities. The authors in \cite{minwalla2} solve the Einstein-dilaton system with negative cosmological constant in which the boundary value of the dilaton field is slowly varying with arbitrary large amplitude. The 
field theory dual of the gravity system becomes a certain fluid dynamics satisfying Navier Stokes equations with dilaton dependent forcing term. We first treat the problem in a naive derivative
expansion.  In this expansion, only the dilaton can have $O(1)$
changes, as in \cite{Awad:2009bh}.  The changes of the metric and
gauge fields are due to the backreaction of the dilaton, and therefore
suppressed by powers of the frequency. In \cite{Awad:2009bh} (as in
\cite{minwalla1,minwalla2}), the equations which determine the
corrections to the fields at any given order $n$ are linear and
contain fields of order $(n-1)$. As expected, we find that the
derivative expansion breaks down and solutions are singular at the
horizon. 

We then explore if the modified low energy limit can tame these
divergences, keeping the variation of the dilaton field $O(1)$.  We
find that the effect of backreaction cannot be ignored in solving for the
corrections to the dilaton and other fields even in the lowest order,
even though the backreaction is small.  This is because the nonlocal
change of the radial coordinate typically implies that radial
derivatives are {\em large} (in terms of the parameter of the
derivative expansion) rather than $O(1)$, thus making the effect of
the backreaction large.

This motivates us to consider the problem for the case where the
dilaton has both small frequency as well as a small amplitude. We calculate the bulk dilaton and its backreaction to the metric and gauge fields to the leading order in amplitude. We find
that in this case the modified low frequency expansion of \cite{Hong1}
and the matching procedure indeed leads to bulk solutions which are
smooth everywhere. This setup is similar to
\cite{Hong1,Edalati1,Edalati2}. 
In these papers, the linearized
problem was solved for fields which also depend on the spatial
coordinates on the boundary, but the backreaction of the fields were
not calculated. In our problem there is no such spatial
dependence; the change of the metric and the gauge fields are entirely
due to the backreaction of the dilaton. A consistent treatment of the backreaction however require us to go to higher orders in the frequency(in some case up to the fourth order). At the same time, we need to go to the second order in the amplitude of the boundary dilaton. The fact that this scheme
works to lowest non-trivial order in the presence of back-reaction indicates that
there is a systematic double expansion in the frequency and the
amplitude which leads to non-singular solutions.

\section{Charged Black Brane with Dilaton Field}
\label{Charged Black Brane with Real Massless Scalar Field}
In this section, we define our model to address the problem. 
We consider 4-dimensional Maxwell-Gravity theory with time dependent dilaton. The solutions are constructed order by order in small frequency in the perturbation theory. It turns out that leading corrections in the perturbation theory are divergent as they approach the black brane horizon in the extremal limit. 

\subsection{Derivative Expansion}
\label{Derivative Expansion}
A consistently truncated theory  from $M$-theory with $S^{7}$ compactification \cite{Duff1} motivates us to consider Einstein-Maxwell-dilaton theory with negative cosmological constant,
\begin{equation}
S=\frac{2}{\kappa^2_{4}} \int d^{4}x \sqrt{-g} \left( \frac{1}{4}R -\frac{1}{4} F_{MN}F^{MN}+\frac{3}{2L^{2}} -\frac{1}{8} \partial_{M} \phi \partial^{M} \phi\right),
\end{equation}
where $\kappa_{4}$ is the gravitational constant. Indices $M$,$N$.. run from 0 to 3, and $F_{MN}$ is field strength from $U(1)$ gauge field $A_{M}$. 
We choose units with $L=1$. The equations of motion are
\begin{eqnarray}
\label{Wequation_in_scalar_field}
W_{MN}&\equiv&R_{MN}+3g_{MN}-2F_{MP}F_{N}^{P}+\frac{1}{2}g_{MN}F_{PQ}F^{PQ} \\ \nonumber
 &-&\frac{1}{2}\partial_{M} {\phi} \partial_{N} \phi=0, \\
\label{Yequation_in_scalar_field}
Y^{N}&\equiv&\nabla_{M}F^{MN}=0, \\
\label{Xequation_in_scalar_field}
X &\equiv& \nabla^2 \phi=0,
\end{eqnarray}
where $\nabla$ denotes a covariant derivative with bulk metric $g_{MN}$.
A charged black brane solution of these equations of motion in Eddington-Finkelstein coordinates with constant dilaton is
\begin{eqnarray}
ds^2 &=&2dvdr-U_{0}(r)dv^2+r^2 dx^{i}dx^{i},  \\
A&=&\rho \left( \frac{1}{r_{0}}-\frac{1}{r} \right) dv, \\
\phi&=&{\rm const},
\end{eqnarray}
where $U_{0}(r)=r^2+\frac{\rho^2}{r^2}-\frac{2\epsilon}{r}$ and $r_{0}$ is the outer horizon of the black brane, the largest root of $U_{0}(r_{0})=0$. $v$ is the ingoing null coordinate which is time coordinate in the $AdS_{4}$ boundary. $\rho$ and $\epsilon$ are charge density and energy density of the black brane respectively. 

Let us consider time dependent dilaton which is slowly varying compared to $r_{0}$, $\phi=\phi_{(0)}(v)$. More precisely, the dialton has a form of
\begin{equation}
\label{parametric_form_of_zeroth_dilaton_field}
\phi_{(0)}(v)=f(\frac{\varepsilon v}{r_{0}}),
\end{equation}
where $\varepsilon$ is dimensionless small parameter. The function $f$ satisfies
\begin{equation}
\label{derivative_argument_parametric_form_of_zeroth_dilaton_field}
f^{\prime}(\frac{\varepsilon v}{r_{0}})\sim O(1),
\end{equation}
where prime denotes derivative with respect to its argument. The derivative of the dilaton with time is suppressed by $\varepsilon$.
\begin{equation}
\frac{d\phi_{0}(v)}{dv}=\frac{\varepsilon}{r_{0}} f^{\prime}(\frac{\varepsilon v}{r_{0}}) \sim \frac{\varepsilon}{r_{0}} \sim O(\varepsilon).
\end{equation}
$\phi_{(0)}(v)$ is obviously not a solution of the equations of motions. To solve the dilaton equation perturbatively, we add correction terms. The dilaton is expanded as
\begin{equation}
\label{Derivative_expnasion_scalar_field_ansatz}
\phi(r,v)=\phi_{(0)}(v)+\phi_{(1)}(r,v)+\phi_{(2)}(r,v)...,
\end{equation}
where $\phi_{(0)}$ is zeroth order in $\varepsilon$, and $\phi_{(1)}$ is 1st order in $\varepsilon$ and so on. The dilaton solution can be calculated perturbatively order by order in $\varepsilon$, which becomes the expansion parameter of the perturbation theory.
We promote the energy density and the charge density to be functions of time as
\begin{eqnarray}
\label{C_definition}
\rho(v)&=&\rho_{0}+r^{2}_{0} C(v), \\
\label{E_definition}
\epsilon(v)&=&\epsilon_{0}+r^{3}_{0} E(v), 
\end{eqnarray}
where $\rho_{0}$ and $\epsilon_{0}$ are constants. For further convenience, we set 
$C(-\infty)=E(-\infty)=0$ as initial conditions. We expand $C(v)$ and $E(v)$ as
\begin{eqnarray}
C(v)&=& C^{(0)}(v)+C^{(1)}(v)+C^{(2)}(v)... \\
E(v)&=& E^{(0)}(v)+E^{(1)}(v)+E^{(2)}(v)...
\end{eqnarray}
The dilaton equation up to first order in $\varepsilon$ becomes \begin{equation}
\label{outer_Scalar_First_order_Equation}
0=\partial_{r} \left( r^2 U(r,v) \partial_{r} \phi_{(1)}(r,v) \right) + 2r\partial_{v} \phi_{(0)}(v),
\end{equation}
where $U(r,v)=r^2+\frac{(\rho_{0}+r^{2}_{0}C(v))^2}{r^2}-\frac{2(\epsilon_{0}+r^{3}_{0}E(v))}{r}$. As we will show below (See Eq(\ref{so-called constraint equations}) and Eq(\ref{so-called constraint equations 2})), 
$C(v)$ and $E(v)$ are higher order in $\varepsilon$.
The first order correction to the dilaton is given by
\begin{equation}
\label{1st_order_scalar_solution_in_outer_region}
\phi_{(1)}(r,v)=\int^{r} \frac{r^{2}_{0}\Lambda_{1}(v)-r^2}{r^2 U_{0}(r)} (\partial_{v}\phi_{(0)}(v)) + \Lambda_{2}(v),
\end{equation}
where $U_{0}(r)=r^2 +\frac{\rho^2_{0}}{r^2}-\frac{2\epsilon_{0}}{r}$ and $\Lambda_{1}(v)$ and $\Lambda_{2}(v)$ are integration constants which are to be determined by boundary conditions that we demand. The regularity condition at the black brane horizon requires $\Lambda_{1}(v)=1$. Moreover, we want a specific boundary condition that as $r \rightarrow \infty$, $\phi(r,v)=\phi_{(0)}(v)$. $\Lambda_{2}(v)$ is determined by this boundary condition. 

We pause here to demonstrate the relationship of the tube-wise solution with derivative expansion. 
Consider the congruence of null geodesics emanating from $AdS_{4}$ boundary and tubes which are centered along the null geodesics. The set up is spatially homogeneous, so we classify the tubes by $v$. Without loss of generality, we set $v=0$ for every individual tube. We expand $\phi_{(0)}(v)$ in the neighborhoods of $v=0$ as
\begin{equation}
\phi_{(0)}(v)=\phi_{(0)}(0)+\varepsilon v\partial_{v}\phi_{(0)}(0)+\frac{1}{2}\varepsilon^{2} v^{2}\partial^{2}_{v}\phi_{(0)}(0)+...
\end{equation}
The charge density and energy density can be expanded as 
\begin{eqnarray}
\rho(v)&=&\rho(0)+\varepsilon v\partial_{v}\rho(0)+\frac{1}{2}\varepsilon^{2} v^{2}\partial^{2}_{v}\rho(0)..., \\
\epsilon(v)&=&\epsilon(0)+\varepsilon v\partial_{v}\epsilon(0)+\frac{1}{2}\varepsilon^{2} v^{2}\partial^{2}_{v}\epsilon(0)...
\end{eqnarray}
We add correction terms to the dialton field as
\begin{equation}
\phi(r)=\phi_{(0)}+\varepsilon \phi_{(1)}(r)+\varepsilon^{2} \phi_{(2)}(r)...
\end{equation}
We omit $v$-dependence in the correction terms because the differential operator acting on these becomes ultra-local as argued below Eq(\ref{ultra-local operator}). Plugging these into the dialton equation and evaluating it up to first order in $\varepsilon$ at $v=0$, we obtain
\begin{equation}
0=\partial_{r} \left( r^2 U(r) \partial_{r} \phi_{(1)}(r) \right) + 2r\partial_{v} \phi_{(0)}(0),
\end{equation}
where $U(r)=r^2+\frac{\rho^2(0)}{r^2}-\frac{2\epsilon(0)}{r}$. The solution of this equation becomes
\begin{equation}
\label{solution_of_dialton_with_navie_derivative exapnsion}
\phi_{(1)}(r)=\int^{r} \frac{r^{2}_{0}\Lambda_{1}(0)-r^2}{r^2 U(r)} (\partial_{v}\phi_{(0)}(0)) + \Lambda_{2}(0),
\end{equation}
where $\Lambda_{1}(0)=1$ for the regularity of the solution $\phi_{(1)}(r)$ at the horizon and $\Lambda_{2}(0)$ is determined by the same boundary condition of our solution at $AdS_{4}$ boundary. To get global solution, we should patch every local solution. Even if the way of getting solution is different, Eq(\ref{1st_order_scalar_solution_in_outer_region}) and Eq(\ref{solution_of_dialton_with_navie_derivative exapnsion}) have the same form at least in the first order in small frequency expansion. This is also true for its back reactions.

Back reactions are obtained perturbatively order by order in $\varepsilon$ with gauge field and metric being expanded as
\begin{eqnarray}
g_{MN}&=&g^{(0)}_{MN}+g^{(1)}_{MN}+g^{(2)}_{MN}...\\
A_{M}&=&A^{(0)}_{M}+A^{(1)}_{M}+A^{(2)}_{M}...
\end{eqnarray}
For leading order correction, we try following form of metric and gauge field solutions:
\begin{eqnarray}
\label{zeroth_order_metric_and_form_filed_equation}
ds^2 &=& 2dvdr-U(r,v)dv^2+r^2 dx^{i}dx^{i} + \frac{k(r,v)}{r^2}dv^2-2h(r,v)drdv, \\ \nonumber
A &=& \rho(v)(\frac{1}{r_{0}(v)}-\frac{1}{r})dv  + a(r,v)dv,
\end{eqnarray}
where $k(r,v)$, $h(r,v)$, and $a(r,v)$ are leading order corrections in the perturbation theory.
Details of equations of motions and calculations of the solutions are in Appendix \ref{Leading Corrections of the Model}. We briefly list the leading back reactions. The leading corrections to gauge field and metric are
\begin{eqnarray}
g^{(1)}_{MN}&=&A^{(1)}_{M}=0, \\
\label{1st_order_scalar_solution_in_outer_region_and_its_backreaction_to_h}
h(r,v) &\equiv& -g^{(2)}_{rv}=-\frac{1}{4} (\partial_{v}\phi_{(0)}(v))^2 \int ^{r}  r^{\prime} \left( \frac{r_{0}^2-r^{\prime 2}}{r^{\prime 2}U_{0}(r^{\prime})} \right)^2 dr^{\prime} + \bar{h}_{1}(v), \\
\label{1st_order_scalar_solution_in_outer_region_and_its_backreaction_to_k}
k(r,v) &\equiv& r^{2}g^{(2)}_{vv}=-\frac{r^{2}}{2}U_{0}(r) (\partial_{v}\phi_{(0)})^2 \int ^{r}  r^{\prime} \left( \frac{r_{0}^2-r^{\prime 2}}{r^{\prime 2}U_{0}(r^{\prime})} \right)^2 dr^{\prime} \\ \nonumber
&+& \frac{r}{4} (\partial_{v}\phi_{(0)}(v))^2  \int ^{r}  \frac{(r_{0}^2-r^{\prime 2})^2}{r^{\prime 2}U_{0}(r^{\prime})}  dr^{\prime} + r \bar{k}_{1}(v)-2\rho_{0} \bar{a}_{1}(v) + 2r^{2}U_{0}(r) \bar{h}_{1}(v), \\
\label{1st_order_scalar_solution_in_outer_region_and_its_backreaction_to_a}
a(r,v) &\equiv& A^{(2)}_{v}=\frac{\rho_{0}}{4} (\partial_{v}\phi_{(0)}(v))^2 \int ^{r}  \frac{dr^{\prime}}{r^{\prime 2}} \int ^{r^{\prime}}  r^{\prime\prime} \left( \frac{r_{0}^2-r^{\prime\prime 2}}{r^{\prime\prime 2}U_{0}(r^{\prime\prime})} \right)^2 dr^{\prime\prime}  - \frac{\bar{a}_{1}(v)-\rho_{0} \bar{h}_{1}(v)}{r} \\ \nonumber
&+& \bar{a}_{2}(v).
\end{eqnarray}
$\bar{h}_{1}(v)$, $\bar{a}_{1}(v)$, $\bar{a}_{2}(v)$ and $\bar{k}_{1}(v)$ are integration constants. They are determined by specific boundary conditions that we demand. At the black brane horizon, these solutions are already regular by choosing $\Lambda_{1}(v)=1$. For the boundary condition at $r=\infty$ we demand that each leading correction of the perturbation theory behaves as
\begin{eqnarray}
\label{Outer_region_Boundary_condiTion}
h(r,v) &\sim& O(r^{0}), \\ \nonumber
k(r,v) &\sim& O(r^{3}), \\ \nonumber
a(r,v) &\sim& O(r^{-2}).
\end{eqnarray}
The motivation behind these boundary conditions is that there are no non-normalizable modes which deform the boundary metric, chemical potential or charge density \cite{minwalla1,minwalla2,Hansen1,Nabamita1}. 

There are the constraint equations which are certain combinations of the equations of back reactions. They are in fact the equations of dual fluid dynamics \cite{minwalla2}.
\begin{eqnarray}
\label{so-called constraint equations}
C^{(0)}(v)&=&C^{(1)}(v)=E^{(0)}(v)=0, \\ 
\label{so-called constraint equations 2}
\dot{E}^{(1)}(v)&=& \frac{1}{4r_{0}} (\partial_{v}\phi_{(0)}(v))^2 .
\end{eqnarray}
The other components of gauge field and metric are trivial.

\subsection{Divergences of the Leading Order Corrections in the Extremal Limit}
\label{Divergences of the Leading Order Perturvation in the Extremal Limit}
The regularity of the solution that we impose for each perturbative correction in the previous section breaks down in the background of extremal black brane.
In Appendix \ref{Outer Solution in Extremal Limit}, we derive the near horizon behavior of leading corrections of the dilaton and its back reactions in the extremal limit by expansion in $u-1$, where $u$ is a rescaled radial coordinate, $u \equiv \frac{r}{r_{0}}$. To be more general, we keep $\Lambda_{1}(v)$ to be arbitrary in Eq(\ref{Near_Black_brane_HiriZon_expansion_of_the _scalar field}). As shown above, $\Lambda_{1}(v)=1$ ensures regularity at the horizon for non-extremal black brane. However in the extremal limit, all the corrections, Eq(\ref{1st_order_scalar_solution_in_outer_region}),Eq(\ref{1st_order_scalar_solution_in_outer_region_and_its_backreaction_to_h}),Eq(\ref{1st_order_scalar_solution_in_outer_region_and_its_backreaction_to_k}) and Eq(\ref{1st_order_scalar_solution_in_outer_region_and_its_backreaction_to_a}) have singularities at the horizon. For example,

\begin{equation}
\label{final_expression_of_outer_expansion_of_dilaton}
\phi(u,v) = \phi_{(0)}
- \frac{\partial_{v}  \phi_{(0)}(v)}{3 r_{0}}ln(u-1) + O(1),
\end{equation}
as $u \rightarrow 1$.
The near horizon expansion of the back reactions are also leading to divergences in physical quantities like curvature invariants and field strengths.
The $u \rightarrow 1$ behavior of the leading correction to the gauge field is given by
\begin{equation}
\label{final_expression_of_outer_expansion_of_a}
a(u,v) = -\frac{\sqrt{3}}{108}\frac{(\partial_{v}\phi_{(0)}(v))^2}{r_{0}}   \left(3ln(u-1)-2(u-1)ln(u-1)+O(u-1)^2 \right) + O(1).
\end{equation}
The first two terms can cause singularity in the field strength, $F_{rv}$. 
The divergences of metric corrections are 
\begin{eqnarray}
\label{final_expression_of_outer_expansion_of_h}
h(u,v)&=&  -\frac{1}{4}\frac{(\partial_{v}\phi_{(0)}(v))^2}{r^{2}_{0}} \left( -\frac{1}{9(u-1)} + \frac{2}{27}ln(u-1) + O(1)\right), \\ 
\label{final_expression_of_outer_expansion_of_k}
k(u,v)&=& -\frac{1}{4}{r^{2}_{0}} (\partial_{v}\phi_{(0)}(v))^{2} \left( \frac{8}{9} (u-1)^2 ln(u-1) + O(1)\right).
\end{eqnarray}
In particular the term multiplying $(u-1)^2 ln(u-1)$ in $k(u,v)$ can possibly cause singularities in curvature invariants.

\subsection{Divergence Resolution (The Main Result)}
\label{Divergence Resolution (The Main Result)}
In this subsection we briefly discuss the main result of this paper without much technical details. 
To deal with divergences discussed in Sec.\ref{Divergences of the Leading Order Perturvation in the Extremal Limit}, we follow Ref.\cite{Hong1} and divide the radial coordinate into two regions.
\begin{eqnarray}
\label{the radial variable change}
{\rm \ Inner\ Region}&:& u-1=\frac{\nu}{\xi} {\rm\ \ for\ \ } \delta < \xi < \infty, \\ \nonumber
{\rm \ Outer\ Region}&:& \frac{\nu}{\delta} < u-1,
\end{eqnarray}
with a certain scaling limit,
\begin{eqnarray}
\label{scaling_limit}
\nu \rightarrow 0,{\rm\ } \xi={\rm finite},{\rm\ } \delta \rightarrow 0, {\ \rm \ and\ \ } \frac{\nu}{\delta} \rightarrow 0.
\end{eqnarray}
where $\nu$ is frequency of the fields in the perturbation theory. Note that switching the radial variable $u$ to $\xi$ is a non-local transformation. For this transformation, we need to evaluate our equations in the frequency space.
We use $\xi$ as a radial coordinate for the inner region and $u$ as that for the outer region. We also define overlapping region (or matching region)as a region that $\xi \sim \delta$. The black brane horizon is located at $\xi=\infty$. As $\xi \rightarrow 0$, we approach the overlapping region by the scaling limit(\ref{scaling_limit}). The solutions listed in Sec.\ref{Derivative Expansion} 
are the outer region solutions. The expansions of the dilaton and its back reactions as $u \rightarrow 1$ in Sec.\ref{Divergences of the Leading Order Perturvation in the Extremal Limit} are therefore the fields in the overlapping region.
With the scaling limit we define a new perturbation theory in small frequency in the inner region. If the inner region solutions are
\begin{itemize}
\item Regular at the black brane horizon,
\item Smoothly connected to the outer region solutions in the overlapping region,
\end{itemize}
the solutions are regular everywhere. 

The dilaton equation in the extremal background is 
\begin{equation}
\nabla^{2} \phi(u,v)=\frac{1}{u^2} \left( \partial_{u}(u^2 U(u,v) \partial_{u} \phi(u,v)) + \frac{1}{r_{0}}\partial_{u} (u^2 \partial_{v} \phi(u,v)) + \frac{u^2}{r_{0}} \partial_{u} \partial_{v} \phi(u,v) \right)=0,
\end{equation}
where again we define dimensionless radial coordinate, $u \equiv \frac{r}{r_0}$. The metric factor $U(u,v)=\frac{(u-1)^2}{u^2} (u^2 +2u +3) + \frac{2\sqrt{3}C(v)+C^{2}(v)}{u^2}-\frac{2E(v)}{u}$. More explicitly, the equation has a form of 
\begin{eqnarray}
\label{More_ExpliciTe_Form_of_the_most_general_scalaR_field}
 0&=&\partial_{u} \left( (u-1)^2 (u^2 +2u +3) \partial_{u} \phi(u,v) \right) + \frac{1}{r_{0}}\partial_{u} (u^2 \partial_{v} \phi(u,v)) + \frac{u^2}{r_{0}} \partial_{u} \partial_{v} \phi(u,v) \\ \nonumber
&-& 2E(v) \partial_{u} \left( u \partial_{u} \phi(u,v) \right).
\end{eqnarray}
In this equation, we see that the term multiplying the energy density $E(v)$ is of $O(\varepsilon^2)$ whereas the terms proportional to $\partial^{2}_{u}\phi(u,v)$ is of $O(\varepsilon)$, using the constraint equation(\ref{so-called constraint equations 2})(which shows $E(v) \sim O(\varepsilon)$) and the dilaton solution(\ref{1st_order_scalar_solution_in_outer_region}). Then, it may appear that the last term in Eq(\ref{More_ExpliciTe_Form_of_the_most_general_scalaR_field}) can be ignored for obtaining the first order solution in $\varepsilon$. However, this is no longer true when we switch the radial variable $u$ to $\xi$. 
The momentum $\nu$ which appears in Eq(\ref{the radial variable change}) is effectively proportional to $\varepsilon$. This is because the dilaton field is localized around $\nu \sim \varepsilon$ in the momentum space(See the discussion in the beginning of Sec.\ref{Divergence Resolution of Massless Scalar Field}). In $\xi$ coordinate, each $u$-derivative in Eq(\ref{More_ExpliciTe_Form_of_the_most_general_scalaR_field}) produces extra factor of $\frac{1}{\varepsilon}$. Therefore, the first term in Eq(\ref{More_ExpliciTe_Form_of_the_most_general_scalaR_field}) becomes the higher order in $\varepsilon$ than
the term multiplying the energy density in $\xi$ coordinate. This means that this later term cannot be ignored any longer.
We have not yet been able to solve this type of equation. In the following we will choose a regime where the amplitude of the dilaton field is small. This will allow us to ignore the last term in Eq(\ref{More_ExpliciTe_Form_of_the_most_general_scalaR_field}), but retain the essential feature of the problem.

The inner region solutions that we solve are completely agree with above two conditions for the entire solutions to be regular everywhere. 
To be precise, we briefly list our inner region solutions. The inner region solution of the dilaton has a form of
\begin{eqnarray}
\label{the_final_result_of_dialton}
\phi_{(in)\nu}(\xi)&=&\phi^{(0)}_{\nu}+ \nu \left( A^{(1)}_{\nu}-\frac{i\phi^{(0)}_{\nu}}{3r_{0}} e^{\frac{i}{3r_{0}}\xi}E_{1}(\frac{i}{3r_{0}}\xi) \right) + \nu^{2} A^{(2)}_{\nu} 
\\ \nonumber
&-& \frac{i r_{0}\nu^2}{2} \left( \int^{\xi}_{\infty}\frac{g_{(2)\nu}(\xi^{\prime})}{\xi^{\prime2}} d \xi^{\prime} - e^{\frac{i}{3r_{0}}\xi} \int^{\xi}_{\infty} \frac{g_{(2)\nu}(\xi^{\prime})}{\xi^{\prime2}} e^{-\frac{i}{3r_{0}}\xi^{{\prime}}}  d \xi^{\prime} \right)+O(\nu)^{3} ,
\end{eqnarray}
where 
\begin{equation}
g_{(2)\nu}(\xi)=\frac{2iA^{(1)}_{\nu}}{r_{0}}-\frac{4i\phi^{(0)}_{\nu}}{3r_{0}\xi}+\frac{2i\phi^{(0)}_{\nu}}{3r_{0}\xi}\left( 1-\frac{4i\xi}{3r_{0}} \right)e^{\frac{i\xi}{3r_{0}}}E_{2}(\frac{i\xi}{3r_{0}}).
\end{equation}
$E_{k}(x) \equiv \int^{\infty}_{1}\frac{e^{-xt}}{t^{k}}dt$ is The Integral Exponential Function, where $k$ is an integer. The numerical coefficients, $A^{(1)}_{\nu}$, $A^{(2)}_{\nu}$ and so on, are determined by matching with the outer region dilaton solution(\ref{final_expression_of_outer_expansion_of_dilaton}) in the overlapping region, more explicitly with Eq(\ref{Near_Black_brane_HiriZon_expansion_of_the _scalar field}). The inner region solutions are developed in the frequency space. Therefore, for matching we need to take the outer region solution to momentum space by Fourier transformation defined in Eq(\ref{define Fourier transformations_scalar}). The precise expression of the outer region solution of the dilaton in frequency space is in Sec.\ref{outer Solution of dialton}. We have a precise form of $A^{(1)}_{\nu}$ as
\begin{equation}
\label{A1 the constant determined}
A^{(1)}_{\nu}=\frac{i \phi^{(0)}_{\nu}}{36 r_{0}} \left( 6\sqrt{2}tan^{-1}(\sqrt{2})-6ln(6) - \frac{6\pi}{\sqrt{2}} \right) +\frac{i\gamma}{3r_{0}}\phi^{(0)}_{\nu}+\frac{i\phi^{(0)}_{\nu}}{3r_{0}}ln\left( \frac{i\nu}{3r_{0}} \right),
\end{equation}
where $\gamma=0.57721..$ is Euler's constant. 
For the regularity at the horizon and the matching with the outer region dilaton solution, only ingoing waves at the black brane horizon are allowed in Eq(\ref{the_final_result_of_dialton}). As a homogeneous solution of the dilaton equation we could add outgoing waves as $B^{(n)}_{\nu}e^{\frac{i\xi}{3r_{0}}}$ to Eq(\ref{the_final_result_of_dialton}), where $B^{(n)}_{\nu}$ is a constant of order $n$ in $\varepsilon$. The matching condition forces $B^{(0)}_{\nu}=0$. For $n > 1$, it turns out that $B^{(n)}_{\nu}$ spoils the regularity of the dilaton solution in $n+1$th order in $\epsilon$. Thus, we naturally impose ingoing boundary condition at the horizon for the smooth dilaton field. 
The other coefficients are obtained by matching with higher order solutions in the outer region.
Near horizon, the inner solutions behave as
\begin{eqnarray}
\phi_{(in)\nu}(\xi)&\sim&\phi^{(0)}_{\nu}+ \nu \left( A^{(1)}_{\nu} -\frac{\phi^{(0)}_{\nu}}{\xi} 
+...  \right)  + \nu^{2}\left( A^{(2)}_{\nu} - \frac{A^{(1)}_{\nu}}{\xi}  \right.\\ \nonumber
 &+& \frac{\phi^{(0)}_{\nu}-3ir_{0}A^{(1)}_{\nu}}{\xi^{2}} +\left....\right) +...,
\end{eqnarray}
which is manifestly regular as $\xi \rightarrow \infty$. 

We evaluate back reactions from the dilaton up to the second order in small frequency as
\begin{eqnarray}
\label{final form of h}
h_{(in)\nu}(\xi)&=&h^{(1)}_{(in)\nu}(\xi)+\nu^{2}\bar{H}^{(2)}_{\nu}+\nu h^{(2)}_{(in)\nu}(\xi)..., \\
\label{final form of a}
a_{(in)\nu}(\xi)&=&\nu^{2}\bar{A}^{(2)}_{\nu}+\nu^2 \tilde{A}^{(2)}_{\nu}+\sqrt{3}r_{0}\nu\int^{\xi}_{\infty}\frac{dy}{y^{2}}h^{(1)}_{(in)\nu}(y)-\sqrt{3}r_{0}\nu^{3}\frac{\bar{H}^{(2)}_{\nu}}{\xi}+\nu^3 \bar{A}^{(3)}_{\nu}+\nu^3 \tilde{A}^{(3)}_{\nu} \\ \nonumber 
&+&\sqrt{3}r_{0}\nu^{2}\int^{\xi}_{\infty}\frac{dy}{y^{2}} \left(  h^{(2)}_{(in)\nu}(y) -\frac{2}{y}h^{(1)}_{(in)\nu}(y)\right)..., \\ 
\label{final form of k}
k_{(in)\nu}(\xi)&=&\nu^{2}\bar{K}^{(2)}_{\nu}+\nu^{3}\left( \bar{K}^{(3)}_{\nu}+\frac{\bar{K}^{(2)}_{\nu}}{\xi} \right)+6r^{4}_{0}\nu^{2}\left( \frac{h^{(1)}_{(in)\nu}(\xi)}{\xi^2} - 2\int^{\xi}_{\infty} \frac{dy}{y^{3}}h^{(1)}_{(in)\nu}(y) \right) \\ \nonumber
&+& \nu^3 \tilde{K}^{(3)}_{\nu}+ \nu^{4} \left( \bar{K}^{(4)}_{\nu} +\frac{\bar{K}^{(3)}_{\nu}}{\xi}\right) +\nu^{4} \left( \tilde{K}^{(4)}_{\nu} +\frac{\tilde{K}^{(3)}_{\nu}}{\xi}\right)+12\nu^{4}r^{4}_{0}\frac{\bar{H}^{(2)}_{\nu}}{\xi^{2}} \\ \nonumber
&-&2r^{4}_{0}\nu^{3}\left( 6\int^{\xi}_{\infty}\frac{dy}{y^{3}}h^{(2)}_{(in)\nu}(y)+3\int^{\xi}_{\infty}\frac{dy}{y^{4}}h^{(1)}_{(in)\nu}(y) -3\frac{h^{(2)}_{(in)\nu}(\xi)}{\xi^{2}} - 2\frac{h^{(1)}_{(in)\nu}(\xi)}{\xi^{3}}\right) \\ \nonumber
&+& 12r^{4}_{0}\nu^{3}\int^{\xi}_{\infty}\frac{dy}{y^{2}}\int^{y}_{\infty}\frac{dz}{z^{3}}h^{(1)}_{(in)\nu}(z)...,
\end{eqnarray}
where
\begin{eqnarray}
h^{(1)}_{(in)\nu}(\xi)&=&\nu \tilde{H}^{(1)}_{\nu}-\frac{1}{36r^{2}_{0}}\int^{\infty}_{-\infty}d \omega \phi^{(0)}_{\omega}\phi^{(0)}_{\nu-\omega} \frac{\omega(\nu-\omega)}{\nu} \int^{\xi}_{\infty}dy \left( 1 -\frac{2i\omega y}{3r_{0}\nu} e^{\frac{i \omega y}{3r_{0}\nu}}E_{1}(\frac{i \omega y}{3r_{0}\nu}) \right. \\ \nonumber
&-&\left.\frac{\omega(\nu-\omega)y^{2}}{9r^{2}_{0}\nu^{2}}e^{\frac{i}{3r_{0}}y}E_{1}(\frac{i \omega y}{3r_{0}\nu})E_{1}(\frac{i (\nu-\omega) y}{3r_{0}\nu}) \right), \\
h^{(2)}_{(in)\nu}(\xi)&=&\nu \tilde{H}^{(2)}_{\nu}-\frac{1}{36r^{2}_{0}}\int^{\infty}_{-\infty}d \omega \phi^{(0)}_{\omega}\phi^{(0)}_{\nu-\omega} \frac{\omega(\nu-\omega)}{\nu} \int^{\xi}_{\infty}\frac{dy}{y} \left( 1 -\frac{2i\omega y}{3r_{0}\nu} e^{\frac{i \omega y}{3r_{0}\nu}}E_{1}(\frac{i \omega y}{3r_{0}\nu}) \right. \\ \nonumber
&-&\left.\frac{\omega(\nu-\omega)y^{2}}{9r^{2}_{0}\nu^{2}}e^{\frac{i}{3r_{0}}y}E_{1}(\frac{i \omega y}{3r_{0}\nu})E_{1}(\frac{i (\nu-\omega) y}{3r_{0}\nu}) \right) \\ \nonumber
&-&\frac{i}{36r_{0}}\int^{\infty}_{-\infty} d\omega \phi^{(0)}_{\nu-\omega} \frac{\omega^{3}(\nu-\omega)^{2}}{\nu^{4}}\int^{\xi}_{\infty} y^{2}dy \left( \frac{\nu}{(\nu-\omega)y} e^{\frac{i\omega y}{3 r_{0} \nu}} \right. \\ \nonumber
&-& \left.\frac{i}{3r_{0}}e^{\frac{i}{3r_{0}}y} E_{1}(\frac{i (\nu-\omega) y}{3r_{0}\nu}) \right) \int^{\frac{\omega}{\nu}y}_{\infty} dz \frac{g_{(2)\omega}(z)}{z^2}e^{-\frac{i}{3r_{0}}z}.
\end{eqnarray}
$\bar{H}^{(2)}_{\nu}$, $\bar{A}^{(2)}_{\nu}$, $\bar{A}^{(3)}_{\nu}$, $\bar{K}^{(2)}_{\nu}$, $\bar{K}^{(3)}_{\nu}$ and $\bar{K}^{(4)}_{\nu}$ are integration constants which are determined by matching with the outer region solutions. 
Eq(\ref{final form of h}), Eq(\ref{final form of a})  and Eq(\ref{final form of k}) are smoothly connected to frequency space expressions of the outer region solutions of Eq(\ref{final_expression_of_outer_expansion_of_h}), Eq(\ref{final_expression_of_outer_expansion_of_a}) and Eq(\ref{final_expression_of_outer_expansion_of_k}) respectively. Details are discussed in Sec.\ref{Matching inner back reaction}.
We determine some of the integration constants in the inner region solutions with this as
\begin{eqnarray}
\bar{H}^{(2)}_{\nu}&=&\frac{1}{864r^{2}_{0}}\int^{\infty}_{-\infty} d\omega \frac{\omega (\nu-\omega)}{\nu^{2}}\phi^{(0)}_{\omega}\phi^{(0)}_{\nu-\omega} \left( 8+\frac{13\sqrt{2}\pi}{2}- 13\sqrt{2}tan^{-1}(\sqrt{2})-8ln(6) \right.\\ \nonumber
&+&\left. 16ln(\nu)\right), \\
\bar{A}^{(2)}_{\nu}&=&-\frac{\sqrt{3}}{288}\int^{\infty}_{-\infty} d\omega \frac{\omega (\nu-\omega)}{\nu^{2}}\phi^{(0)}_{\omega}\phi^{(0)}_{\nu-\omega} \left( -10+2\sqrt{2}tan^{-1}(\sqrt{2})-\sqrt{2}\pi+4ln(6) \right. \\ \nonumber
&-& \left.8ln(\nu)\right), \\
\bar{K}^{(2)}_{\nu}&=& \frac{1}{4}r^{2}_{0}\int^{\infty}_{-\infty}d\omega \frac{\omega(\nu-\omega)}{\nu^{2}}\phi^{(0)}_{\omega}\phi^{(0)}_{\nu-\omega}\left( \sqrt{2}tan^{-1}(\sqrt{2})-1 \right)+\frac{r_{0}\bar{k}^{\nu}_{1}}{\nu^2}.
\end{eqnarray}
$\bar{A}^{(3)}_{\nu}$, $\bar{K}^{(3)}_{\nu}$ and $\bar{K}^{(4)}_{\nu}$ are determined by higher orders in the outer region solutions. $\bar{k}^{\nu}_{1}$ is the Fourier transform of $\bar{k}_{1}(v)$. $\tilde{H}^{(1)}_{\nu}$, $\tilde{H}^{(2)}_{\nu}$, $\tilde{A}^{(2)}_{\nu}$, $\tilde{A}^{(3)}_{\nu}$, $\tilde{K}^{(3)}_{\nu}$ and $\tilde{K}^{(4)}_{\nu}$ are finite numerical constants, which cannot be obtained analytically(In principle we can. See Sec.\ref{Inner Solution in Divergence Resolution of Back Reacted Metric and Gauge Field} for the discussion about these constants).
The near horizon behaviors of the back reactions are given by 
\begin{eqnarray}
{\ \ }h_{(in)\nu}(\xi) &\sim& \nu\tilde{H}^{(1)}_{\nu}+\nu^{2}\bar{H}^{(2)}_{\nu}+\nu^{2}\tilde{H}^{(2)}_{\nu}...+O(\frac{1}{\xi}), {\ \ } \\ \nonumber
a_{(in)\nu}(\xi) &\sim& \nu^{2}\bar{A}^{(2)}_{\nu}+\nu^{2}\tilde{A}^{(2)}_{\nu}+\nu^{3}\bar{A}^{(3)}_{\nu}+\nu^{3}\tilde{A}^{(3)}_{\nu}...+O(\frac{1}{\xi}), \\ \nonumber
{\rm and\ \ } k_{(in)\nu}(\xi) &\sim& \nu^{2}\bar{K}^{(2)}_{\nu}+\nu^{3}\bar{K}^{(3)}_{\nu}+\nu^{3}\tilde{K}^{(3)}_{\nu}+\nu^{4}\bar{K}^{(4)}_{\nu}+\nu^{4}\tilde{K}^{(4)}_{\nu}...+O(\frac{1}{\xi}).
\end{eqnarray}
Thus the solutions are regular everywhere. 

Finally, the constraint equations in the inner region turns out to be the same with those in the outer region(\ref{so-called constraint equations}), which have forms of
\begin{eqnarray}
C^{(0)}_{\nu}&=&C^{(1)}_{\nu}=E^{(0)}_{\nu}=0, \\ \nonumber
\nu^2 E^{(1)}_{\nu}&=&-\frac{1}{4i r_{0}}\int^{\infty}_{-\infty} d\omega \omega(\nu-\omega)\phi^{(0)}_{\omega}\phi^{(0)}_{\nu-\omega} {\rm \ \ and\ so\ on...},
\end{eqnarray}
in the frequency space.

\section{Divergence Resolution of Dilaton Field}
\label{Divergence Resolution of Massless Scalar Field}
In this section, we solve the dialton equation in the inner region. The scaling limit(\ref{scaling_limit}) cannot be applied to Eq(\ref{More_ExpliciTe_Form_of_the_most_general_scalaR_field}) because switching radial variable $u$ to $\xi$ is non-local transformation.
To deal with this, we need to rewrite the equation in the frequency space by Fourier transformation as
\begin{equation}
\label{define Fourier transformations_scalar}
\phi(u,v)=\int^{\infty}_{-\infty} e^{i \omega v} \phi_{\omega}(u) d \omega,
\end{equation} 
where $\phi_{\omega}(u)$ is a localized and normalizable function in frequency space. 
For example, we can choose $\phi_{\omega}(u)$ to be
\begin{equation}
\label{specific form of phi}
\phi_{\omega}(u) \sim e^{-\frac{\omega^2}{\varepsilon^2}}e^{-\frac{\varepsilon^2}{\omega^2}}f(\frac{\omega}{\varepsilon})g_{\omega}(u),
\end{equation}
where $g_{\omega}(u)$ is a function carrying radial variable $u$ and $f(\frac{\omega}{\varepsilon})$ is arbitrary $O(1)$ function of in frequency space. Then, $\phi_{\omega}(u)$ is suppressed as $\omega$ approach either zero or $\infty$ by the exponential factors in it. Consequently, this function is extremely localized around $\omega=\pm \varepsilon$ and normalizable. Using the argument in Appendix \ref{Counting Order in Perturbation Theory}, one can show that Fourier transformation of Eq(\ref{specific form of phi}) becomes a form as Eq(\ref{parametric_form_of_zeroth_dilaton_field}) in the frequency space. This shows that the properties of Eq(\ref{specific form of phi}) is consistent with those of $\phi_{(0)}(v)$ introduced in Sec.\ref{Derivative Expansion}.
The dilaton equation in momentum space can be read off by acting an integral operator $\frac{1}{2\pi}\int^{\infty}_{-\infty} e^{-i \nu v} dv$ on Eq(\ref{More_ExpliciTe_Form_of_the_most_general_scalaR_field}). The equation in the momentum space has a form of
\begin{eqnarray}
\label{More_ExpliciTe_Form_of_the_most_general_scalaR_field in momentum space}
0&=& \partial_{u} \left((u-1)^2 (u^2 +2u +3) \partial_{u} \phi_{\nu}(u) \right)+ \frac{i \nu}{r_{0}}\partial_{u} (u^2  \phi_{\nu}(u)) + \frac{i \nu u^2}{r_{0}} \partial_{u} \phi_{\nu}(u) \\ \nonumber
&-&2 \int^{\infty}_{-\infty} \partial_{u} \left( u \partial_{u} \phi_{\omega}(u) \right) E_{\nu-\omega} d\omega,
\end{eqnarray}
where $E_{\nu-\omega}$ is Fourier transfor of $E(v)$ as defined in Eq(\ref{Fourier_transformation_of_C_and_E}).
In the inner and outer region, we expand the dilaton field as
\begin{eqnarray}
\label{dialton inner region expansion ansatz}
{\rm Inner{\rm\ } Region :\ \ }
\phi_{(in)\nu}(\xi)&=&\phi^{(0)}_{(in)\nu}(\xi)+\nu \phi^{(1)}_{(in)\nu}(\xi)+\nu^{2} \phi^{(2)}_{(in)\nu}(\xi)+...,\\
{\rm Outer{\rm\ } Region :\ \ \ \ \ }
\phi_{\nu}(u)&=&\phi^{(0)}_{\nu}+\nu \phi^{(1)}_{\nu}(u)+\nu^{2} \phi^{(2)}_{\nu}(u)+...,
\end{eqnarray}
respectively.
To avoid confusion, we do not tag outer region solutions with ``$(out)$".

\subsection{Inner Solution} 
As discussed in Sec.\ref{Divergence Resolution (The Main Result)}, we solve linear dilaton equation by ignoring the last term in Eq(\ref{More_ExpliciTe_Form_of_the_most_general_scalaR_field in momentum space}). 
Switching radial variable $u$ to $\xi$, the dilaton equation in the inner region becomes
\begin{equation}
\xi^2 \partial_{\xi} \left( (6+\frac{4\nu}{\xi}+\frac{\nu^2}{\xi^2})\partial_{\xi} \phi_{(in)\nu}(\xi) \right) -\frac{2i\xi^2}{r_{0}}(1+\frac{\nu}{\xi})^{2} \partial_{\xi}\phi_{(in)\nu}(\xi) + \frac{2i\xi^2}{r_{0}} (\frac{\nu}{\xi^2}+\frac{\nu^2}{\xi^3}) \phi_{(in)\nu}(\xi)=0.
\end{equation} 
The zeroth order equation in $\nu$ is 
\begin{equation}
6\xi^2 \partial^{2}_{\xi} \phi^{(0)}_{(in)\nu}(\xi) - \frac{2i\xi^2}{r_{0}} \partial_{\xi} \phi^{(0)}_{(in)\nu}(\xi)=0.
\end{equation}
This equation gives two linearly independent solutions,
\begin{equation}
\phi^{(0)}_{(in)\nu}=A^{(0)}_{\nu}+B^{(0)}_{\nu}e^{\frac{i}{3r_{0}}\xi},
\end{equation}
where $A^{(0)}_{\nu}$ is purely incoming wave and the term multiplying $B^{(0)}_{\nu}$ is purely outgoing wave at the extremal black brane horizon: $\xi=\infty$. The subscript ``$\nu$" denotes that the integration constants depend on frequency. These two independent solutions are regular at the horizon up to the zeroth order in $\nu$. First order equation in $\nu$ is
\begin{equation}
6 \xi^2 \partial^{2}_{\xi} \phi^{(1)}_{(in)\nu}(\xi)-\frac{2i\xi^2}{r_{0}} \partial_{\xi} \phi^{(1)}_{(in)\nu}(\xi) + \frac{2i}{r_{0}} \left( A^{(0)}_{\nu} +B^{(0)}_{\nu}e^{\frac{i}{3r_{0}}\xi} \right)=0.
\end{equation}
The solution of this equation is
\begin{equation}
\label{the most general scalar solution}
\phi^{(1)}_{(in)\nu}=-\frac{iA^{(0)}_{\nu}}{3r_{0}} \int^{\xi}_{\infty} e^{\frac{i}{3r_{0}}\xi^{\prime}} d\xi^{\prime} \left( \int^{\xi^{\prime}}_{\infty} \frac{e^{-\frac{i}{3r_{0}}\xi^{\prime\prime}}}{\xi^{\prime\prime2}} d\xi^{\prime\prime} + B^{(1)}_{\nu} \right) +A^{(1)}_{\nu}+ \frac{iB^{(0)}_{\nu}}{3r_{0}}\int^{\xi}_{\infty}\frac{e^{\frac{i}{3r_{0}}\xi^{\prime}}}{\xi^{\prime}}d\xi^{\prime},
\end{equation}
where $A^{(1)}_{\nu}$ and $B^{(1)}_{\nu}$ are integration constants, which are corresponding to incoming and outgoing waves at the horizon respectively. 
Using
\begin{equation}
\int^{x} e^{i \alpha x^{\prime}} dx^{\prime} \int^{x^{\prime}} \frac{e^{-i \alpha x^{\prime\prime}}}{x^{\prime\prime2}}dx^{\prime\prime} = e^{i\alpha x}E_{1}(i\alpha x),
\end{equation}
Eq(\ref{the most general scalar solution}) becomes
\begin{equation}
\label{first_order_solution_in_complex_scalar_field}
\phi^{(1)}_{(in)\nu}(\xi)=A^{(1)}_{\nu}-A^{(0)}_{\nu}B^{(1)}_{\nu}e^{\frac{i\xi}{3r_{0}}} -\frac{iA^{(0)}_{\nu}}{3r_{0}} e^{\frac{i}{3r_{0}}\xi}E_{1}(\frac{i}{3r_{0}}\xi) - \frac{iB^{(0)}_{\nu}}{3r_{0}}E_{1}(-\frac{i}{3r_{0}}\xi)
\end{equation}
The first and the third terms are incoming waves at the horizon, whereas the others are outgoing waves. Let us discuss the asymptotic form of the solution.   
For large $y$, $E_{1}(y)$ is expanded as
\begin{equation}
\label{E_1_identity_for_large_y}
E_{1}(y)= \frac{e^{-y}}{y}\sum_{n=0}^{\infty}\frac{(-1)^{n}(n)! }{y^{n}},
\end{equation}
where $y$ is pure imaginary number, so $e^{-y}$ term is bounded. It is manifest that $E_{1}(y)$ is regular as $y$ approaches infinity. In the case that $y$ goes to zero, the function $E_{1}(y)$ becomes divergent. Let us argue what the leading divergence is. To see the leading divergence, we calculate following object:
\begin{equation}
\lim_{y \rightarrow 0}\frac{E_{1}(y)}{ln(y)}= \lim_{y \rightarrow 0}\frac{\frac{d E_{1}(y)}{dy}}{\frac{d ln(y)}{dy}}=-1,
\end{equation}
where we have used Hospital's theorem for the first equality. 
Then, the leading divergent term is logarithmic. 
Consequently, this solution is regular at the horizon and divergent logarithmically near the matching region. 

\subsection{Outer Solution}
\label{outer Solution of dialton}
In Appendix \ref{Outer Solution in Extremal Limit}, we evaluate the dilaton in outer region as $u \rightarrow 1$ in the extremal limit. The solution in momentum space is given by
\begin{equation}
\label{Matching_scalar_equation}
\phi_{\nu}(u) \equiv  (\phi^{(0)}_{\nu}+\nu\phi^{(1)}_{\nu}(u))=\phi^{(0)}_{\nu}+ \frac{i\nu \phi^{(0)}_{\nu}}{r_{0}} \int ^{u} \frac{\Lambda_{1}-u^2}{u^2U_{0}(u)} du +\Lambda_{2},
\end{equation}
where $\Lambda_{1}$ and $\Lambda_{2}$ come from the Fourier transforms of the integration constants $\Lambda_{1}(v)$ and $\Lambda_{2}(v)$. They depend on $\nu$. Near matching region expansion of Eq(\ref{Matching_scalar_equation}) is
\begin{eqnarray}
\label{Near_Black_brane_horizon_expansion_of_the _scalar_field_in_momentum_space}
\phi_{\nu}(u)&=&\phi^{(0)}_{\nu}+\frac{i \nu  \phi^{(0)}_{\nu}}{6(u-1)r_{0}} (1 - \Lambda_{1}) - \frac{i\nu  \phi^{(0)}_{\nu}}{9 r_{0}}ln(u-1) \left( 2 + \Lambda_{1} \right) \\ \nonumber
&+& \frac{i\nu \phi^{(0)}_{\nu}}{36 r_{0}} \left( (\Lambda_{1}-7)\sqrt{2}tan^{-1}(\sqrt{2}) + 2(\Lambda_{1}+2)ln(6) + \frac{\pi}{\sqrt{2}}(7- \Lambda_{1}) \right) \\ \nonumber
&+& \frac{i\nu \phi^{(0)}_{\nu}}{18r_{0}}(u-1)+O(u-1)^2.
\end{eqnarray}

\subsection{Matching}
\label{Matching}
For the inner solution to match the outer one, we need to switch the radial coordinate $\xi$ to $u$ near matching region. The matching region is defined as a region with $\frac{\nu}{u-1} \sim \delta$. The scaling limit(\ref{scaling_limit}) shows that $\frac{\nu}{u-1}$ can become a small expansion parameter near matching region. The outer region solution is perturbative solution order by order in $\nu$. It is justified that one can do series expansion of $\phi_{(in)\nu}$ in $\nu$ for matching the outer solution. The inner region solution up to the leading order correction in $\nu$, Eq(\ref{first_order_solution_in_complex_scalar_field}), in the radial variable $u$ is given by
\begin{eqnarray}
\phi_{(in)\nu}(u)&=&A^{(0)}_{\nu}+B^{(0)}_{\nu}e^{\frac{i\nu}{3r_{0}(u-1)}}+\nu \left( A^{(1)}_{\nu}-A^{(0)}_{\nu}B^{(1)}_{\nu}e^{\frac{i\nu}{3r_{0}(u-1)}} \right.\\ \nonumber
&-&\left.\frac{iA^{(0)}_{\nu}}{3r_{0}}e^{\frac{i\nu}{3r_{0}(u-1)}}E_{1}(\frac{i\nu}{3r_{0}(u-1)})-\frac{iB^{(0)}_{\nu}}{3r_{0}}E_{1}(\frac{-i\nu}{3r_{0}(u-1)})\right).
\end{eqnarray}
Using asymptotic expansion of $E_{1}(y)$ for small $y$ as
\begin{equation}
\label{E_1_identity_for_small_y}
E_{1}(y)=-\gamma-ln(y)-\sum^{\infty}_{n=1}\frac{(-1)^{n}y^{n}}{n n!},
\end{equation}
we expand the inner region solution in terms of $\frac{\nu}{u-1}$. The expansion has a form of
\begin{eqnarray}
\label{matching inner to outer of dilaton}
\phi_{(in)\nu}(u)&=&A^{(0)}_{\nu}+B^{(0)}_{\nu}+\nu\left( \frac{iB^{(0)}_{\nu}}{3r_{0}(u-1)}-\frac{i}{3r_{0}}(A^{(0)}_{\nu}+B^{(0)}_{\nu})ln(u-1)+A^{(1)}_{\nu} \right.\\ \nonumber
&-&\left.A^{(0)}_{\nu}B^{(1)}_{\nu}+\frac{i\gamma}{3r_{0}}(A^{(0)}_{\nu}+B^{(0)}_{\nu})+\frac{iA^{(0)}_{\nu}}{3r_{0}}ln\left( \frac{i\nu}{3r_{0}} \right)+\frac{iB^{(0)}_{\nu}}{3r_{0}}ln\left( -\frac{i\nu}{3r_{0}} \right) \right)\\ \nonumber
&+&O(\frac{\nu}{u-1})^{2}
\end{eqnarray}
We compare the asymptotes of $\phi_{(in)\nu}(u)$ with Eq(\ref{Near_Black_brane_horizon_expansion_of_the _scalar_field_in_momentum_space}) to determine each coefficient in it. At the zeroth order in $\nu$,
\begin{equation}
\label{Zeroth_order_MaTching_with_A0_and_B0}
A^{(0)}_{\nu}+B^{(0)}_{\nu}=\phi^{(0)}_{\nu}.
\end{equation}
At the first order,
\begin{eqnarray}
\label{First_order_MaTching_with_A0_and_B0}
A^{(0)}_{\nu}+B^{(0)}_{\nu}&=&\frac{2+\Lambda_{1}}{3}\phi^{(0)}_{\nu}, \\
\label{First_order_MaTching_with_B0}
B^{(0)}_{\nu}&=&\frac{1-\Lambda_{1}}{2}\phi^{(0)}_{\nu}, \\
\label{First_order_MaTching_with_A1_and_B1}
A^{(0)}_{\nu}B^{(1)}_{\nu}-A^{(1)}_{\nu}&=&\frac{i \phi^{(0)}_{\nu}}{36 r_{0}} \left( (\Lambda_{1}-7)\sqrt{2}tan^{-1}(\sqrt{2}) + 2(\Lambda_{1}+2)ln(6) + \frac{\pi}{\sqrt{2}}(7- \Lambda_{1}) \right) \\ \nonumber
&-&\frac{i\gamma}{3r_{0}}(A^{(0)}_{\nu}+B^{(0)}_{\nu})-\frac{iA^{(0)}_{\nu}}{3r_{0}}ln\left( \frac{i\nu}{3r_{0}} \right)-\frac{iB^{(0)}_{\nu}}{3r_{0}}ln\left( -\frac{i\nu}{3r_{0}} \right).
\end{eqnarray}
Eq(\ref{Zeroth_order_MaTching_with_A0_and_B0}), Eq(\ref{First_order_MaTching_with_A0_and_B0}), Eq(\ref{First_order_MaTching_with_B0}) and Eq(\ref{First_order_MaTching_with_A1_and_B1}) provide
\begin{equation}
\label{LambdaAB}
\Lambda_{1}=1, {\ \ } A^{(0)}_{\nu}=\phi^{(0)}_{\nu}, {\ \ } B^{(0)}_{\nu}=0,
\end{equation}
and
\begin{equation}
\label{A1}
A^{(1)}_{\nu}-A^{(0)}_{\nu}B^{(1)}_{\nu}=\frac{i \phi^{(0)}_{\nu}}{6 r_{0}} \left( \sqrt{2}tan^{-1}(\sqrt{2}) - ln(6) - \frac{\pi}{\sqrt{2}} \right)+\frac{i\gamma \phi^{(0)}_{\nu}}{3r_{0}}+\frac{i\phi^{(0)}_{\nu}}{3r_{0}}ln\left( \frac{i\nu}{3r_{0}} \right).
\end{equation}

\subsection{More on Dilaton Solution in Inner Region}
As is clear from Eq(\ref{matching inner to outer of dilaton}), the lowest order solution in the inner region, expressed in terms of coordinate $u$ contains all power of $\nu$. This is true for all higher order corrections to the inner region solution as well. More precisely, $\nu^{n}\phi^{(n)}_{(in)\nu}(\xi)$, expressed in terms of $u$ will have terms of $O(\nu^m)$ with $m \geqslant 1$. These terms are crucial in ensuring a smooth matching with the outer region solution. For example, in Eq(\ref{Near_Black_brane_horizon_expansion_of_the _scalar_field_in_momentum_space}) there is a term $\sim \nu \phi^{(0)}_{\nu}(u-1)$- but such a term is not present in $\nu \phi^{(1)}_{(in)\nu}(u)$. We now show that such a term is actually present in $\nu^2 \phi^{(2)}_{(in)\nu}(u)$ with precisely the correct coefficient.
To see this, let us evaluate the second order correction of dialton field.
The second order equation is
\begin{equation}
6\xi^2 \partial^{2}_{\xi} \phi^{(2)}_{(in)\nu}(\xi) - \frac{2i\xi^2}{r_{0}} \partial_{\xi} \phi^{(2)}_{(in)\nu}(\xi) + g_{(2)\nu}(\xi)=0,
\end{equation}
where $g_{(2)\nu}(\xi)$ is 
\begin{equation}
g_{(2)\nu}(\xi)=4\xi \phi^{(1)\prime\prime}_{(in)\nu}(\xi)-4(1+\frac{i\xi}{r_{0}})\phi^{(1)\prime}_{(in)\nu}(\xi) +\frac{2i}{r_{0}} \phi^{(1)}_{(in)\nu}(\xi) + \frac{2i}{\xi r_{0}} \phi^{(0)}_{(in)\nu}.
\end{equation}
The prime indicates derivative respect to $\xi$. 
The solution of this equation is 
\begin{equation}
\label{The_most_general_form_of_2nd_order_equation_of_scalar}
\phi^{(2)}_{(in)\nu}(\xi)= A^{(2)}_{\nu} +\frac{i B^{(2)}_{\nu} r_{0}}{2} e^{\frac{i}{3r_{0}}\xi} - \frac{i r_{0}}{2} \left( \int^{\xi}_{\infty}\frac{g_{(2)\nu}(\xi^{\prime})}{\xi^{\prime2}} d \xi^{\prime} - e^{\frac{i}{3r_{0}}\xi} \int^{\xi}_{\infty} \frac{g_{(2)\nu}(\xi^{\prime})}{\xi^{\prime2}} e^{-\frac{i}{3r_{0}}\xi^{{\prime}}}  d \xi^{\prime} \right),
\end{equation}
where we set $B^{(2)}_{\nu}=0$ because it cause logarithmic divergence at the horizon in the third order in $\nu$. Near horizon, $\phi^{(2)}_{(in)\nu}(\xi)$ is expanded as
\begin{equation}
\label{The_near_horizon_form_of_2nd_order_equation_of_scalar}
\phi^{(2)}_{(in)\nu}(\xi) = \sum^{\infty}_{n=0} \frac{\alpha^{n}_{\nu}}{\xi^{n}} + \phi^{(0)}_{\nu}B^{(1)}_{\nu} \left( -\frac{4i}{9r_{0}}ln\xi +\frac{1}{\xi}+... \right),
\end{equation}
where $\alpha^{0}_{\nu}=A^{(2)}_{\nu}$, $\alpha^{1}_{\nu}=-A^{(1)}_{\nu}$, $\alpha^{2}_{\nu}=\phi^{(0)}_{\nu}-3i r_{0}A^{(1)}_{\nu}$ and so on.
Regularity condition of $\phi^{(2)}_{(in)\nu}(\xi)$ at the horizon forces $B^{(1)}_{\nu}=0$. Only incoming wave is allowed in $\phi^{(1)}_{(in)\nu}(\xi)$ too. Near the matching region, we switch the radial variable $\xi$ to $u$ for matching with the outer solution. Defining a new integral variable $y$ as $\xi^{\prime} \equiv \frac{\nu}{y-1}$, Eq(\ref{The_most_general_form_of_2nd_order_equation_of_scalar}) becomes
\begin{equation}
\phi^{(2)}_{(in)\nu}(u) = A^{(2)}_{\nu} 
+ \frac{i r_{0}}{2\nu} \left( \int^{u} g_{(2)\nu}(\frac{\nu}{y-1}) dy - e^{\frac{i\nu}{3r_{0}(u-1)}} \int^{u} g_{(2)\nu}(\frac{\nu}{y-1}) e^{-\frac{i\nu}{3r_{0}(y-1)}}  d y \right).
\end{equation}
For matching, we expand $\phi^{(2)}_{(in)\nu}(u)$ in $\nu$ as
\begin{eqnarray}
\label{The_near_matching_form_of_2nd_order_equation_of_scalar}
\phi^{(2)}_{(in)\nu}(u)  &=& \frac{i\phi^{(0)}_{\nu}(u-1)}{18 r_{0} \nu} +\sum^{\infty}_{j=0}\frac{\nu^{j}}{(u-1)^{j}}\left(\beta^{(2)}_{j\nu}+\beta^{(2)\prime}_{j\nu}ln(u-1)+\beta^{(2)\prime\prime}_{j\nu}\left(ln(u-1)\right)^{2}\right),
\end{eqnarray}
where $\beta^{(2)}_{j\nu}$,$\beta^{(2)\prime}_{j\nu}$ and $\beta^{(2)\prime\prime}_{j\nu}$ are $O(1)$ constants, some of which are given by  $\beta^{(2)\prime}_{0}=\frac{6iA^{(1)}_{\nu}r_{0}+\phi^{(0)}_{\nu}}{18r^{2}_{0}}$, $\beta^{(2)\prime\prime}_{0}=-\frac{\phi^{(0)}_{\nu}}{54r^{2}_{0}}$ and so on. The first term in Eq(\ref{The_near_matching_form_of_2nd_order_equation_of_scalar}) is proportional to $\nu$,
which matches the last term in Eq(\ref{Near_Black_brane_horizon_expansion_of_the _scalar_field_in_momentum_space}) precisely. We expect that similar mechanism ensures matching of the higher order terms.

\section{Divergence Resolution of Back Reacted Metric and Gauge Field}
\label{Divergence Resolution of Back Reacted Metric and Gauge Field}
In this section, we extend our discussion into back reactions. In Appendix.\ref{Master Equations in Extremal Backgroud}, we obtain the equations of the back reactions without ignoring time derivatives. 
These equations are the starting point for our discussion.
\subsection{Inner Solution}
\label{Inner Solution in Divergence Resolution of Back Reacted Metric and Gauge Field}

We begin with Eq(\ref{Einstein equations in momentum space_Wrr_in_appendix}). To solve this equation in the inner region, we need to substitute the inner region solution of the dilaton field into it. To do this, we take the inner region dilaton solution back to the outer region: $\phi_{(in)\omega}(\xi) \rightarrow \phi_{(in)\omega}(\frac{\omega}{u-1})$ and plug it into Eq(\ref{Einstein equations in momentum space_Wrr_in_appendix}). With this, Eq(\ref{Einstein equations in momentum space_Wrr_in_appendix}) becomes
\begin{equation}
\partial_{u}h_{\nu}(u)=-\frac{u}{4}\int^{\infty}_{-\infty} d \omega \partial_{u}\phi_{(in)\nu-\omega}(\frac{\nu-\omega}{u-1})\partial_{u}\phi_{(in)\omega}(\frac{\omega}{u-1}).
\end{equation}
The $u$-derivative acting on the dilaton can be switched to derivative with respect to its argument as
\begin{eqnarray}
\partial_{u}\phi_{(in)\omega}(\frac{\omega}{u-1})=-\frac{\omega}{(u-1)^2}\phi^{\prime}_{(in)\omega}(\frac{\omega}{u-1}).
\end{eqnarray}
After this, we replace the radial coordinate $u$ with $\xi$. 
Then, Eq(\ref{Einstein equations in momentum space_Wrr_in_appendix}) becomes
\begin{equation}
\label{Einstein equations in momentum space_Wrr_inner_EQUATIONS}
h_{(in)\nu}^{\prime}(\xi)=\frac{\xi^2 (1+\frac{\nu}{\xi})}{4\nu^3} \left( \int^{\infty}_{-\infty} d\omega \omega(\nu-\omega) \phi^{\prime}_{(in)\omega}(\frac{\omega}{\nu}\xi) \phi^{\prime}_{(in)\nu-\omega}(\frac{\nu-\omega}{\nu}\xi)\right),
\end{equation}
where the prime on the dilaton denotes derivative with respect to its argument, whereas the prime on $h_{(in)\nu}(\xi)$ does derivative with respect to $\xi$. 
Plugging the dilaton expansion(\ref{dialton inner region expansion ansatz}) into Eq(\ref{Einstein equations in momentum space_Wrr_inner_EQUATIONS}), it becomes
\begin{eqnarray}
\label{h mometum in dynamical equaiton}
h_{(in)\nu}^{\prime}(\xi)&=&\frac{\xi^2}{4 \nu^3}(1+\frac{\nu}{\xi})\int^{\infty}_{-\infty} d \omega \omega^{2}(\nu-\omega)^{2} \left( \phi^{(1)\prime}_{(in)\omega}(\frac{\omega}{\nu}\xi) \phi^{(1)\prime}_{(in)\nu-\omega}(\frac{\nu-\omega}{\nu}\xi) \right. \\ \nonumber
&+& \omega \phi^{(2)\prime}_{(in)\omega}(\frac{\omega}{\nu}\xi) \phi^{(1)\prime}_{(in)\nu-\omega}(\frac{\nu-\omega}{\nu}\xi)  +(\nu-\omega)\phi^{(1)\prime}_{(in)\omega}(\frac{\omega}{\nu}\xi) \phi^{(2)\prime}_{(in)\nu-\omega}(\frac{\nu-\omega}{\nu}\xi) \\ \nonumber
&+&\left....\right).
\end{eqnarray} 
$h_{(in)\nu}(\xi)$ is also localized function around $\nu \sim \pm \varepsilon$ in frequency space as the dilaton field (See the discussion below Eq(\ref{specific form of phi}) in Sec.\ref{Divergence Resolution of Massless Scalar Field}). Then, we expand $h_{(in)\nu}(\xi)$ as
\begin{equation}
\label{entire_solution_of_H_inner}
h_{(in)\nu}(\xi)=\bar{H}_{\nu}+h^{(1)}_{(in)\nu}(\xi) + \nu h^{(2)}_{(in)\nu}(\xi) +...,
\end{equation}
where $h^{(1)}_{(in)\nu}(\xi)$ is in the first order in small frequency and $\nu h^{(2)}_{(in)\nu}(\xi)$ is in the second order and so on. We expand the other corrections of the back reactions in the same way. $\bar{H}_{\nu}$ is an integration constant which can be expanded as $\bar{H}_{\nu}=\nu\bar{H}^{(1)}_{\nu}+\nu^{2}\bar{H}^{(2)}_{\nu}...$, where $\bar{H}^{(1)}_{\nu}$ and $\bar{H}^{(2)}_{\nu}$... are $O(1)$ constants. 
We note that counting power of $\varepsilon$ of the solutions to show that each solution is in the correct order in small frequency expansion is not manifest in the frequency space. In Appendix \ref{Counting Order in Perturbation Theory}, we discuss details about this power counting by scaling all the frequencies appeared in the solutions with $\varepsilon$.
The solutions up to the second order in small frequency are
\begin{eqnarray}
\label{The solutions up to the first subleading_of_2nd_order_equation_of_H1}
h^{(1)}_{(in)\nu}(\xi)&=& \nu\tilde{H}^{(1)}_{\nu}+\int^{\xi}_{\infty} \frac{\xi^{\prime 2} d\xi^{\prime}}{4 \nu^3} \int^{\infty}_{-\infty} d \omega \omega^{2} (\nu-\omega)^{2} \phi^{(1)\prime}_{(in)\omega}(\frac{\omega}{\nu}\xi^{\prime}) \phi^{(1)\prime}_{(in)\nu-\omega}(\frac{\nu-\omega}{\nu}\xi^{\prime}), \\
\label{The_near_horizon_form_of_2nd_order_equation_of_H2}
h^{(2)}_{(in)\nu}(\xi) &=& \nu\tilde{H}^{(2)}_{\nu}+ \int^{\xi}_{\infty} \frac{\xi^{\prime 2} d\xi^{\prime}}{4 \nu^4} \int^{\infty}_{-\infty} d \omega \omega^{2} (\nu-\omega)^{2} \left( \frac{\nu}{\xi^{\prime}}\phi^{(1)\prime}_{(in)\omega}(\frac{\omega}{\nu}\xi^{\prime}) \phi^{(1)\prime}_{(in)\nu-\omega}(\frac{\nu-\omega}{\nu}\xi^{\prime}) \right. \\ \nonumber
&+& \left. \omega \phi^{(2)\prime}_{(in)\omega}(\frac{\omega}{\nu}\xi^{\prime}) \phi^{(1)\prime}_{(in)\nu-\omega}(\frac{\nu-\omega}{\nu}\xi^{\prime}) +(\nu-\omega)\phi^{(1)\prime}_{(in)\omega}(\frac{\omega}{\nu}\xi^{\prime}) \phi^{(2)\prime}_{(in)\nu-\omega}(\frac{\nu-\omega}{\nu}\xi^{\prime}) \right). \\ \nonumber
\end{eqnarray}
As will be shown below, the terms containing $\xi^{\prime}$-integrations appearing in Eq(\ref{The solutions up to the first subleading_of_2nd_order_equation_of_H1}) and Eq(\ref{The_near_horizon_form_of_2nd_order_equation_of_H2}) vanish as $\xi \rightarrow \infty$. However, the near matching region expansions of those terms present additive constant terms. $\tilde{H}^{(1)}_{\nu}$ and $\tilde{H}^{(2)}_{\nu}$ are $O(1)$ numerical constants which are designed so that they precisely cancel those additive constant terms.
With such a choice of $\tilde{H}^{(1)}_{\nu}$ and $\tilde{H}^{(2)}_{\nu}$, the only constant term in the overlapping region expansion of $h_{(in)\nu}(\xi)$ to match that in the outer region solution $h_{\nu}(u)$ become $\bar{H}_{\nu}$ (See Eq(\ref{entire_solution_of_H_inner}) and Eq(\ref{inner expansion of h in this section})). There will be numerical constants as $\tilde{A}^{(2)}_{\nu}$, $\tilde{A}^{(3)}_{\nu}$, $\tilde{K}^{(3)}_{\nu}$ and $\tilde{K}^{(4)}_{\nu}$ appearing in $a_{(in)\nu}(\xi)$ and $k_{(in)\nu}(\xi)$ to be determined by the same manner. They provide the near matching region expansions of $a_{(in)\nu}(\xi)$ and $k_{(in)\nu}(\xi)$ to be exactly given by Eq(\ref{the inner region expansion in terms of U-1 of A}) and Eq(\ref{the inner region expansion in terms of U-1 of K}). We cannot determine these numerical constants analytically but in principle one can obtain the precise values. 

Let us discuss regularity of the solutions.
$\bar{H}_{\nu}$ is regular everywhere. To see the behaviors of $h^{(1)}_{(in)\nu}(\xi)$ and $h^{(2)}_{(in)\nu}(\xi)$ we expand $\phi^{(1)\prime}_{(in)\omega}(x)$ and $\phi^{(2)\prime}_{(in)\omega}(x)$ in Eq(\ref{The solutions up to the first subleading_of_2nd_order_equation_of_H1}) and Eq(\ref{The_near_horizon_form_of_2nd_order_equation_of_H2}) in the limit of large value of their argument $x$ using that Eq(\ref{first_order_solution_in_complex_scalar_field}), Eq(\ref{E_1_identity_for_large_y}), Eq(\ref{The_near_horizon_form_of_2nd_order_equation_of_scalar}) and $B^{(0)}_{\nu}=B^{(1)}_{\nu}=0$. They are given by
\begin{eqnarray}
\label{phi 1 prime}
\phi^{(1)\prime}_{(in)\omega}(x)&=&-\frac{i\phi^{(0)}_{\omega}}{3r_{0}x}\sum^{\infty}_{n=1}\frac{(-1)^{n}n!}{(\frac{ix}{3r_{0}})^{n}}, \\
\label{phi 2 prime}
\phi^{(2)\prime}_{(in)\omega}(x)&=&-\sum^{\infty}_{n=1}\frac{n \alpha^{n}_{\omega}}{x^{n+1}}.
\end{eqnarray}
Substitution of Eq(\ref{phi 1 prime}) to Eq(\ref{The solutions up to the first subleading_of_2nd_order_equation_of_H1}) provides near horizon expansion of $h^{(1)}_{(in)\nu}(\xi)$ as
\begin{equation}
\label{The_near_horizon_of_first_order_solution_of_H}
h^{(1)}_{(in)\nu}(\xi)=\nu\tilde{H}^{(1)}_{\nu}+\sum^{\infty}_{n,m=1}\frac{A_{mn}}{\xi^{m+n-1}},
\end{equation}
where 
\begin{equation}
\label{Amn equation}
A_{mn}=\frac{1}{4}\frac{m!n!(-1)^{m+n+1}}{m+n-1}\nu^{m+n-1}\left( \frac{i}{3r_{0}} \right)^{2-m-n} \int^{\infty}_{-\infty}\phi^{(0)}_{\omega}\phi^{(0)}_{\nu-\omega}\omega^{1-n}(\nu-\omega)^{1-m}d\omega,
\end{equation}
where $m$ and $n$ are integers. 
We note that the $\omega$-integration in Eq(\ref{Amn equation}) seems to have poles at $\omega=0$ and $\omega=\nu$ for $m+n > 2$ and could lead to an infinite integrand. However, we set $\phi^{(0)}_{\omega} \sim e^{-\frac{\varepsilon^2}{\omega^2}}$ as $\omega \rightarrow 0$ 
as discussed in the beginning of Sec.\ref{Divergence Resolution of Massless Scalar Field} (See Eq(\ref{specific form of phi})). This ensures that the integration is finite. Then, 
near horizon behavior of $h^{(1)}_{(in)\nu}(\xi)$ is given by  $h^{(1)}_{(in)\nu}(\xi) \sim \nu\tilde{H}^{(1)}_{\nu}+ O(\frac{1}{\xi})$.

We obtain near horizon behavior of $h^{(2)}_{(in)\nu}(\xi)$  by plugging Eq(\ref{phi 1 prime}) and Eq(\ref{phi 2 prime}) into Eq(\ref{The_near_horizon_form_of_2nd_order_equation_of_H2}), which is given by
\begin{equation}
\label{The_near_horizon_of_second_order_solution_of_H}
h^{(2)}_{(in)\nu}(\xi)=\nu\tilde{H}^{(2)}_{\nu}+\sum^{\infty}_{n,m=1} \left( \frac{m+n-1}{m+n}\frac{A_{mn}}{\xi^{m+n}}  + \frac{B_{mn}}{\xi^{m+n-1}}\right),
\end{equation}
where
\begin{eqnarray}
\label{Bmn equation}
B_{mn}&=&\frac{1}{4}\frac{m!(n)(-1)^{m+1}}{m+n-1}\nu^{m+n-2}\left( \frac{i}{3r_{0}} \right)^{1-m} \int^{\infty}_{-\infty}\left( \phi^{(0)}_{\nu-\omega}\alpha^{n}_{\omega}\omega^{2-n}(\nu-\omega)^{1-m} \right.\\ \nonumber
&+& \left.\phi^{(0)}_{\omega}\alpha^{n}_{\nu-\omega}\omega^{1-m}(\nu-\omega)^{2-n} \right) d\omega,
\end{eqnarray}

The $\alpha^{n}_{\omega}$ in Eq(\ref{Bmn equation}) are proportional to $\phi^{(0)}_{\omega}$ (See Eq(\ref{LambdaAB}), Eq(\ref{A1}),  Eq(\ref{The_near_horizon_form_of_2nd_order_equation_of_scalar}) and discussion below it). This ensures that the integration in Eq(\ref{Bmn equation}) is also finite. Consequently,  $h^{(2)}_{(in)\nu}(\xi) \sim \nu\tilde{H}^{(2)}_{\nu}+O(\frac{1}{\xi})$ near horizon.

Secondly we solve Eq(\ref{gauge field equations in momentum space_Yr_in_appendix}) which provides solutions of gauge field corrections in the inner region. Switching radial variable $u$ to $\xi$, Eq(\ref{gauge field equations in momentum space_Yr_in_appendix}) becomes
\begin{equation}
\label{inner a equation in cluding c}
a^{\prime}_{(in)\nu}(\xi)=\frac{\nu r_{0}}{\xi^{2}(1+\frac{\nu}{\xi})^{2}} \left( C_{\nu}+\sqrt{3}h_{(in)\nu}(\xi)\right).
\end{equation}
We expand the charge density as $C_{\nu}=C^{(0)}_{\nu}+\nu C^{(1)}_{\nu}+\nu^{2} C^{(2)}_{\nu}...$.
$a_{(in)\nu}(\xi)$ can be expanded as $a_{(in)\nu}(\xi)= \bar{A}_{\nu}+a^{(1)}_{(in)\nu}(\xi) + \nu a^{(2)}_{(in)\nu}(\xi) + \nu^2 a^{(3)}_{(in)\nu}(\xi)...$, where again $\bar{A}_{\nu}$ is an integration constant which is also expanded as $\bar{A}_{\nu}=\nu\bar{A}^{(1)}_{\nu}+\nu^2 \bar{A}^{(2)}_{\nu}...$ 
The solutions up to the third order expansion are given by
\begin{eqnarray}
a^{(1)}_{(in)\nu}(\xi) &=& -r_{0}\nu\frac{C^{(0)}_{\nu}}{\xi}, \\
a^{(2)}_{(in)\nu}(\xi) &=& \nu\tilde{A}^{(2)}_{\nu} + \sqrt{3}r_{0}  \int^{\xi}_{\infty} \frac{d \xi^{\prime}}{\xi^{\prime 2}}h^{(1)}_{(in)\nu}(\xi^{\prime})- \frac{\sqrt{3}r_{0}\nu}{\xi}\bar{H}^{(1)}_{\nu} + r_{0}\nu\left( \frac{C^{(0)}_{\nu}}{\xi^{2}} -\frac{C^{(1)}_{\nu}}{\xi} \right), \\
a^{(3)}_{(in)\nu}(\xi) &=& \nu\tilde{A}^{(3)}_{\nu}+\sqrt{3}r_{0}  \int^{\xi}_{\infty} \frac{d \xi^{\prime}}{\xi^{\prime 2}} \left( h^{(2)}_{(in)\nu}(\xi^{\prime}) -\frac{2}{\xi^{\prime}}h^{(1)}_{(in)\nu}(\xi^{\prime}) \right)+\sqrt{3}r_{0}\nu \left( \frac{\bar{H}^{(1)}_{\nu}}{\xi^{2}}-\frac{\bar{H}^{(2)}_{\nu}}{\xi} \right)\\ \nonumber
&-&r_{0}\nu \left( \frac{C^{(0)}_{\nu}}{\xi^{3}}-\frac{C^{(1)}_{\nu}}{\xi^{2}}+\frac{C^{(2)}_{\nu}}{\xi} \right).
\end{eqnarray}
The reason why we need to obtain $a_{(in)\nu}(\xi)$ up to the third order in the small frequency is that when we plug the $n$th order solution of $h_{(in)\nu}(\xi)$ into Eq(\ref{inner a equation in cluding c}), we get $n+1$th order solution of $a_{(in)\nu}(\xi)$. For the same reason, we need to get the inner solution of $k_{(in)\nu}(\xi)$ up to the fourth order in small frequency.

Let us explore near horizon limit of the solutions. $a^{(1)}_{(in)\nu}(\xi)$ is manifestly regular at the horizon. Using Eq(\ref{The_near_horizon_of_first_order_solution_of_H}) and Eq(\ref{The_near_horizon_of_second_order_solution_of_H}), we evaluate near horizon expansions of $a^{(2)}_{(in)\nu}$ and $a^{(3)}_{(in)\nu}$ which have forms of
\begin{eqnarray}
a^{(2)}_{(in)\nu}&=&  \nu\tilde{A}^{(2)}_{\nu} -\sqrt{3}r_{0} \sum^{\infty}_{m,n=1}\frac{A_{mn}}{m+n}\frac{1}{\xi^{m+n}}- \frac{\sqrt{3}r_{0}\nu}{\xi} \bar{H}^{(1)}_{\nu}+ r_{0}\nu\left( \frac{C^{(0)}_{\nu}}{\xi^{2}} -\frac{C^{(1)}_{\nu}}{\xi} \right) \\ \nonumber &-& \frac{\sqrt{3}r_{0}\nu}{\xi} \tilde{H}^{(1)}_{\nu}, \\
a^{(3)}_{(in)\nu}&=& \nu\tilde{A}^{(3)}_{\nu} +
\sqrt{3}r_{0}  \sum^{\infty}_{m,n=1}\left(\frac{A_{mn}}{m+n}\frac{1}{\xi^{m+n+1}} - \frac{B_{mn}}{m+n}\frac{1}{\xi^{m+n}}\right)+\sqrt{3}r_{0}\nu \frac{\bar{H}^{(1)}_{\nu}}{\xi^2}\\ \nonumber
&-&\sqrt{3}r_{0}\nu\frac{\bar{H}^{(2)}_{\nu}}{\xi}-r_{0}\nu \left( \frac{C^{(0)}}{\xi^{3}}-\frac{C^{(1)}}{\xi^{2}}+\frac{C^{(2)}}{\xi} \right)+\sqrt{3}r_{0}\nu\left(\frac{\tilde{H}^{(1)}_{\nu}}{\xi^2}-\frac{\tilde{H}^{(2)}_{\nu}}{\xi} \right).
\end{eqnarray}
Then, $a_{(in)\nu}(\xi)$ is finite at the horizon.

Combining Eq(\ref{Einstein equations in momentum space_Wii_in_appendix}) and Eq(\ref{gauge field equations in momentum space_Yv_in_appendix}), we get
\begin{equation}
\label{Combining_equation_of_Einstein equations in momentum space_Wii_in_appendix_and_auge field equations in momentum space_Yv_in_appendix}
0=-6(u^{4}-1)h_{\nu}(u)-u(u^{4}-4u+3)h^{\prime}_{\nu}(u) +\frac{1}{r^{4}_{0}} \left( uk^{\prime}_{\nu}(u) - k_{\nu}(u)\right) + 2\sqrt{3}C_{\nu},
\end{equation}
which gives solutions of $k_{(in)\nu}(\xi)$.
Changing radial coordinate $u$ into $\xi$, this equation becomes 
\begin{eqnarray}
\label{inner region k eq}
0&=&\nu^2 \left( 6+ \frac{10\nu}{\xi} + \frac{5\nu^2}{\xi^2} + \frac{\nu^3}{\xi^3}\right)h^{\prime}_{(in)\nu}(\xi) - 6\nu\left( \frac{4\nu}{\xi}+ \frac{6\nu^2}{\xi^2} + \frac{4\nu^3}{\xi^3} + \frac{\nu^4}{\xi^4} \right)h_{(in)\nu}(\xi) \\ \nonumber 
&+&2\sqrt{3}\nu C_{\nu} -\frac{1}{r^{4}_{0}}\left( \nu k_{(in)\nu}(\xi)+ (\xi^2 + \nu\xi)k^{\prime}_{(in)\nu}(\xi) \right).
\end{eqnarray}
We expand $k_{(in)\nu}(\xi)$ as $k_{(in)\nu}(\xi)= \bar{K}_{\nu}(\xi)+k^{(1)}_{(in)\nu}(\xi) + \nu k^{(2)}_{(in)\nu}(\xi) + \nu^2 k^{(3)}_{(in)\nu}(\xi)...$, where $\bar{K}_{\nu}(\xi)$ is a homogeneous solution of Eq(\ref{inner region k eq}) which satisfies
\begin{equation}
\label{K-equation}
\nu \bar{K}_{\nu}(\xi)+ (\xi^2 + \nu\xi)\bar{K}^{\prime}_{\nu}(\xi)=0.
\end{equation}
The solution of Eq(\ref{K-equation}) is given by
\begin{equation}
\bar{K}_{\nu}(\xi)=\nu\bar{K}^{(1)}_{\nu}+\nu^{2} \left( \bar{K}^{(2)}_{\nu} +\frac{\bar{K}^{(1)}_{\nu}}{\xi} \right)+ \nu^{3}\left( \bar{K}^{(3)}_{\nu} +\frac{\bar{K}^{(2)}_{\nu}}{\xi} \right)...,
\end{equation}
where $\bar{K}^{(1)}_{\nu}$, $\bar{K}^{(2)}_{\nu}$ and $\bar{K}^{(3)}_{\nu}$... are arbitrary $O(1)$ constants. We solve Eq(\ref{inner region k eq}) up to fourth order in small frequency which are given by
\begin{eqnarray}
\label{the inner region and expanding of solution K1}
k^{(1)}_{(in)\nu}(\xi)&=& -2\sqrt{3}r^{4}_{0}\nu\frac{C^{(0)}_{\nu}}{\xi} \\
k^{(2)}_{(in)\nu}(\xi)&=& -2\sqrt{3}r^{4}_{0}\nu\frac{C^{(1)}_{\nu}}{\xi} \\
k^{(3)}_{(in)\nu}(\xi)&=& \nu\tilde{K}^{(3)}_{\nu} +6r^{4}_{0} \int^{\xi}_{\infty} \frac{d\xi^{\prime}}{\xi^{\prime 2}} \left( h^{(1)\prime}_{(in)\nu}(\xi^{\prime})-\frac{4}{\xi^{\prime}} h^{(1)}_{(in)\nu}(\xi^{\prime})\right)+ 12r^{4}_{0}\nu \frac{\bar{H}^{(1)}_{\nu}}{\xi^{2}} \\ \nonumber
&-&2\sqrt{3}r^{4}_{0}\nu\frac{C^{(2)}_{\nu}}{\xi}, \\
\label{the inner region and expanding of solution K2}
k^{(4)}_{(in)\nu}(\xi) &=& \nu\tilde{K}^{(4)}_{\nu}-6r^{4}_{0}\int^{\xi}_{\infty}\frac{d\xi^{\prime}}{\xi^{\prime 2}} \int^{\xi^{\prime}}_{\infty}\frac{d\xi^{\prime\prime}}{\xi^{\prime\prime 2}} \left( h^{(1)\prime}_{(in)\nu}(\xi^{\prime\prime})-\frac{4}{\xi^{\prime\prime}} h^{(1)}_{(in)\nu}(\xi^{\prime\prime})\right) \\ \nonumber
&-& 2r^{4}_{0}  \int^{\xi}_{\infty} \frac{d\xi^{\prime}}{\xi^{\prime 2}} \left( \frac{12}{\xi^{\prime}}h^{(2)}_{(in)\nu}(\xi^{\prime}) +\frac{6}{\xi^{\prime 2}}h^{(1)}_{(in)\nu}(\xi^{\prime}) -3h^{(2) \prime}_{(in)\nu}(\xi^{\prime}) -\frac{2}{\xi^{\prime}}h^{(1) \prime}_{(in)\nu}(\xi^{\prime}) \right) \\ \nonumber
&+&12r^{4}_{0}\nu\frac{\bar{H}^{(2)}_{\nu}}{\xi^{2}}+8r^{4}_{0}\nu\frac{\bar{H}^{(1)}_{\nu}}{\xi^{3}}-2\sqrt{3}r^{4}_{0}\nu\frac{C^{(3)}_{\nu}}{\xi}+\nu \frac{\tilde{K}^{(3)}_{\nu}}{\xi}.
\end{eqnarray}
$\bar{K}_{\nu}(\xi)$, $k^{(1)}_{(in)\nu}(\xi)$ and $k^{(2)}_{(in)\nu}(\xi)$ are manifestly regular at the horizon. 
We list behaviors of $k^{(3)}_{(in)\nu}(\xi)$ and $k^{(4)}_{(in)\nu}(\xi)$ near horizon:  
\begin{eqnarray}
k^{(3)}_{(in)\nu}(\xi)&=& \nu\tilde{K}^{(3)}_{\nu}+12r^{4}_{0}\nu \frac{\bar{H}^{(1)}_{\nu}+\tilde{H}^{(1)}_{\nu}}{\xi^{2}} -2\sqrt{3}r^{4}_{0}\nu\frac{C^{(2)}_{\nu}}{\xi}+ 6r^4_{0}\sum^{\infty}_{m,n=1}\frac{m+n+3}{m+n+1}\frac{A_{mn}}{\xi^{m+n+1}}, \\
k^{(4)}_{(in)\nu}(\xi)&=&\nu\tilde{K}^{(4)}_{\nu}+12r^{4}_{0}\nu\frac{\bar{H}^{(2)}_{\nu}+\tilde{H}^{(2)}_{\nu}}{\xi^{2}}+8r^{4}_{0}\nu\frac{\bar{H}^{(1)}_{\nu}+\tilde{H}^{(1)}_{\nu}}{\xi^{3}}-2\sqrt{3}r^{4}_{0}\nu\frac{C^{(3)}_{\nu}}{\xi} +\nu \frac{\tilde{K}^{(3)}_{\nu}}{\xi}    \\ \nonumber 
&+&r^4_{0}\sum^{\infty}_{m,n=1}  \frac{6(m+n+3)}{m+n+1} \frac{B_{mn}}{\xi^{m+n+1}} +r^4_{0}\sum^{\infty}_{m,n=1}\frac{A_{mn}}{\xi^{m+n+2}}\left( \frac{10m+10n+2}{m+n+2} \right. \\ \nonumber
&+&  \left. \frac{6(m+n+3)}{(m+n+1)(m+n+2)} + \frac{24(m+n-1)}{(m+n)(m+n+2)} \right)  .
\end{eqnarray}
As $\xi \rightarrow \infty$, $k^{(3)}_{(in)\nu}(\xi) \sim \nu\tilde{K}^{(3)}_{\nu}+ O(\frac{1}{\xi})$ and $k^{(4)}_{(in)\nu}(\xi) \sim \nu\tilde{K}^{(4)}_{\nu}+ O(\frac{1}{\xi})$. Therefore, the inner region solutions are regular solutions.

Finally, we solve Eq(\ref{Master Equations in Extremal Background_W-bar}) to obtain the constraint equations in the inner region, which are given by
\begin{eqnarray}
\label{inner constraint 1}
0&=&E^{(0)}_{\nu}-\sqrt{3}C^{(0)}_{\nu}, \\
\label{inner constraint 2}
0&=&i\nu^2 r_{0} \left( -2\sqrt{3}C^{(1)}_{\nu} + 2E^{(1)}_{\nu} +\frac{\bar{K}^{(1)}_{\nu}}{r^{4}_{0}}\right) + \frac{1}{2}\int^{\infty}_{-\infty} d \omega (\nu-\omega)\omega \phi^{(0)}_{\nu-\omega}\phi^{(0)}_{\omega}, \\
\label{inner constraint 3}
0&=&i\nu^3 r_{0} \left( -2\sqrt{3}C^{(2)}_{\nu} + 2E^{(2)}_{\nu} +\frac{\bar{K}^{(2)}_{\nu}}{r^{4}_{0}}\right) -\int^{\infty}_{-\infty} d \omega (\nu-\omega)\omega^2 \phi^{(0)}_{\nu-\omega} A^{(1)}_{\omega}.
\end{eqnarray}

\subsection{Matching}
\label{Matching inner back reaction}
In this subsection, we show that the inner region solutions solved in the previous subsection are smoothly connected to the outer region solutions in Sec.\ref{Derivative Expansion}. Let us start with $h_{(in)\nu}(\xi)$. Near matching region, we expand Eq(\ref{The solutions up to the first subleading_of_2nd_order_equation_of_H1}) with small $\xi$ using Eq(\ref{first_order_solution_in_complex_scalar_field}) and Eq(\ref{E_1_identity_for_small_y}). 
Integrating by $\xi$, Eq(\ref{The solutions up to the first subleading_of_2nd_order_equation_of_H1}) becomes
\begin{equation}
\label{H1_expansion}
h^{(1)}_{(in)\nu}(\frac{\nu}{u-1})=\frac{1}{4}\left( \frac{i}{3r_{0}} \right)^{2}\int^{\infty}_{-\infty}d\omega \omega(\nu-\omega)\phi^{(0)}_{\omega}\phi^{(0)}_{\nu-\omega}\frac{1}{u-1}+{\rm \ subleading\ terms...},
\end{equation}
where the ``subleading terms" denote terms which are higher order in $\nu$ when $h^{(1)}_{(in)\nu}(\xi)$ is expressed in terms of $u$ (We obtain leading corrections only in the outer region). The same procedure is applied to Eq(\ref{The_near_horizon_form_of_2nd_order_equation_of_H2}). Near the overlapping region, the expansion of $h^{(2)}_{(in)\nu}(u)$ becomes
\begin{equation}
\label{H2_expansion}
\nu h^{(2)}_{(in)\nu}(\frac{\nu}{u-1})=-\frac{1}{6}\left( \frac{i}{3r_{0}} \right)^{2}\int^{\infty}_{-\infty}d\omega \omega(\nu-\omega)\phi^{(0)}_{\omega}\phi^{(0)}_{\nu-\omega}ln\left(\frac{u-1}{\nu}\right)+{\rm \ subleading\ terms...}
\end{equation}
Combining these, we get near matching region expansion of $h_{(in)\nu}(u)$ which is a form of
\begin{equation}
\label{inner expansion of h in this section}
h_{(in)\nu}(u)=\nu \bar{H}^{(1)}_{\nu}+\nu^2 \bar{H}^{(2)}_{\nu}+\frac{1}{4}\left( \frac{i}{3r_{0}} \right)^{2}\int^{\infty}_{-\infty}d\omega \omega(\nu-\omega)\phi^{(0)}_{\omega}\phi^{(0)}_{\nu-\omega}\left( \frac{1}{u-1} -\frac{2}{3}ln(\frac{u-1}{\nu}) \right)+...
\end{equation}
As $u \rightarrow 1$, the outer region solution of $h_{\nu}$ in momentum space is expanded as
\begin{eqnarray}
\label{outer expnasion of h in this section in the section}
h_{\nu}(u)&=&\frac{1}{4}(\frac{i}{3r_{0}})^{2} \int^{\infty}_{-\infty} d\omega \omega (\nu-\omega)\phi^{(0)}_{\omega}\phi^{(0)}_{\nu-\omega} \left( \frac{1}{u-1}-\frac{2}{3}ln(u-1)-\frac{1}{24}\left( 8+\frac{13\sqrt{2}\pi}{2}\right.\right.\\ \nonumber &-&\left.\left. 13\sqrt{2}tan^{-1}(\sqrt{2})-8ln(6) \right) +... \right)
\end{eqnarray}
(See Appendix \ref{Outer Solution in Extremal Limit}).
Consequently, Eq(\ref{inner expansion of h in this section})
completely matches Eq(\ref{outer expnasion of h in this section in the section})
requesting that $\bar{H}^{(1)}_{\nu}=0$, $\nu^{2} \bar{H}^{(2)}_{\nu}=\frac{1}{864r^{2}_{0}}\int^{\infty}_{-\infty} d\omega \omega (\nu-\omega)\phi^{(0)}_{\omega}\phi^{(0)}_{\nu-\omega} \left( 8+\frac{13\sqrt{2}\pi}{2}- 13\sqrt{2}tan^{-1}(\sqrt{2})-8ln(6) +16ln(\nu)\right)$, and so on.

In the overlapping region, $a_{(in)\nu}(u)$ and $k_{(in)\nu}(u)$ are obtained in the same way.  As $\xi$ approaches zero, 
$a_{(in)\nu}(u)$ and $k_{(in)\nu}(u)$ are expanded as
\begin{eqnarray}
\label{the inner region expansion in terms of U-1 of A}
a_{(in)\nu}(u)&=&-r_{0}C^{(0)}_{\nu} \left( (u-1)-(u-1)^2+(u-1)^3... \right) -r_{0}\nu C^{(1)}_{\nu} \left( (u-1)-(u-1)^2... \right) \\ \nonumber
&-&r_{0}\nu^2 C^{(2)}_{\nu}(u-1) ... +\nu\bar{A}^{(1)}_{\nu}+\nu^{2}\bar{A}^{(2)}_{\nu}+\nu^{2}\bar{A}^{(3)}_{\nu}... \\ \nonumber
&-&\sqrt{3}r_{0}\nu \bar{H}^{(1)}_{\nu} \left( (u-1)-(u-1)^2... \right)-\sqrt{3}r_{0}\nu^2 \bar{H}^{(2)}_{\nu}(u-1)...\\ \nonumber
&+&\frac{\sqrt{3}r_{0}}{4}(\frac{i}{3r_{0}})^{2}\int^{\infty}_{-\infty}d\omega \omega (\nu-\omega)\phi^{(0)}_{\omega}\phi^{(0)}_{\nu-\omega} \left(-ln\left(\frac{u-1}{\nu}\right)+ \frac{2}{3}(u-1)ln\left(\frac{u-1}{\nu}\right)\right.\\ \nonumber
&+&\left.\frac{4}{3}(u-1)...\right), \\
\label{the inner region expansion in terms of U-1 of K}
k_{(in)\nu}(u)&=&\nu\bar{K}^{(1)}_{\nu}+\nu(u-1)\bar{K}^{(1)}_{\nu}+\nu^2 \bar{K}^{(2)}_{\nu}+\nu^2 (u-1)\bar{K}^{(2)}_{\nu}+\nu^3 \bar{K}^{(3)}_{\nu} \\ \nonumber
&+&\nu^3 (u-1)\bar{K}^{(3)}_{\nu} +\nu^4 \bar{K}^{(4)}_{\nu}... -2\sqrt{3}r^{4}_{0}(u-1)\left( C^{(0)}_{\nu}+ \nu C^{(1)}_{\nu} +\nu^2 C^{(2)}_{\nu} + \nu^3 C^{(3)}_{\nu}...\right) \\ \nonumber
&+& 4r^{4}_{0}\nu \bar{H}^{(1)}_{\nu}\left( 3(u-1)^2 +2(u-1)^3... \right)+12r^{4}_{0}\nu^2 (u-1)^2 \bar{H}^{(2)}_{\nu}... +\nu^3 \tilde{K}^{3}_{\nu}(u-1)...\\ \nonumber
&+&\frac{1}{4}r^{2}_{0}\int^{\infty}_{-\infty}d\omega \omega (\nu-\omega)\phi^{(0)}_{\omega}\phi^{(0)}_{\nu-\omega} \left( \frac{8}{9}(u-1)^2 ln\left(\frac{u-1}{\nu}\right)-2(u-1)\right. \\ \nonumber
&-&\left. \frac{5}{3}(u-1)^2 ...\right).
\end{eqnarray}
In the overlapping region, the outer region solutions $a_{\nu}(u)$ and $k_{\nu}(u)$ in momentum space are given by
\begin{eqnarray}
\label{the outer region expansion in terms of U-1 of A}
a_{\nu}(u)&=&\frac{\sqrt{3}r_{0}}{4}(\frac{i}{3r_{0}})^{2}\int^{\infty}_{-\infty}d\omega \omega (\nu-\omega)\phi^{(0)}_{\omega}\phi^{(0)}_{\nu-\omega} \left(-ln(u-1)+ \frac{2}{3}(u-1)ln(u-1)\right. \\ \nonumber
&+&\frac{1}{8}\left( -10+2\sqrt{2}tan^{-1}(\sqrt{2})-\sqrt{2}\pi+4ln(6) \right)\\ \nonumber
&+&\frac{1}{24}\left( 40-13\sqrt{2}tan^{-1}(\sqrt{2})-8ln(6)+\frac{13\sqrt{2}\pi}{2} \right)(u-1)\left.+...\right)+O(u-1)^{2}, \\
\label{the outer region expansion in terms of U-1 of K}
k_{\nu}(u)&=& \frac{1}{4}{r^{2}_{0}} \int^{\infty}_{-\infty} d\omega \omega (\nu-\omega)\phi^{(0)}_{\omega}\phi^{(0)}_{\nu-\omega}  \left( \frac{8}{9} (u-1)^2 ln(u-1) -1 + \sqrt{2} tan^{-1} (\sqrt{2}) \right. \\ \nonumber
&+&\left. ( \sqrt{2}tan^{-1}(\sqrt{2})-3)(u-1) + \left( \frac{13\sqrt{2}\pi}{36}-\frac{11}{9}-\frac{4ln(6)}{9} \right.\right. \\ \nonumber
&-&\left.\left. \frac{13tan^{-1}(\sqrt{2})}{9\sqrt{2}} \right)(u-1)^{2}\right)+r_{0} \bar{k}^{\nu}_{1} +r_{0} \bar{k}^{\nu}_{1} (u-1) + O(u-1)^{3},
\end{eqnarray}
where again $\bar{k}^{\nu}_{1}$ is the Fourier transform of $\bar{k}_{1}(v)$.
We compare Eq(\ref{the inner region expansion in terms of U-1 of A}), Eq(\ref{the inner region expansion in terms of U-1 of K}) with Eq(\ref{the outer region expansion in terms of U-1 of A}), Eq(\ref{the outer region expansion in terms of U-1 of K}) respectively to decide that $C^{(0)}_{\nu}=C^{(1)}_{\nu}=C^{(2)}_{\nu}=\bar{A}^{(1)}_{\nu}=\bar{K}^{(1)}_{\nu}=0$,
\begin{eqnarray}
\nonumber
\nu^{2}\bar{A}^{(2)}_{\nu}&=&-\frac{\sqrt{3}}{288}\int^{\infty}_{-\infty} d\omega \omega (\nu-\omega)\phi^{(0)}_{\omega}\phi^{(0)}_{\nu-\omega} \left( -10+2\sqrt{2}tan^{-1}(\sqrt{2})-\sqrt{2}\pi\right. \\ \nonumber
&+&\left. 4ln(6)-8ln(\nu) \right) \\ \nonumber
{\rm  and\ }
\nu^{2}\bar{K}^{(2)}&=& \frac{1}{4}r^{2}_{0}\int^{\infty}_{-\infty}d\omega \omega(\nu-\omega)\phi^{(0)}_{\omega}\phi^{(0)}_{\nu-\omega}\left( \sqrt{2}tan^{-1}(\sqrt{2})-1 \right)+r_0 \bar{k}^{\nu}_{1}.
\end{eqnarray} 
$\bar{A}^{(3)}_{\nu}$, $\bar{K}^{(3)}_{\nu}$ and $\bar{K}^{(4)}_{\nu}$ are determined by matching with higher orders in the outer region solutions. With the coefficients determined in this fashion, two solutions are connected smoothly in the matching region. 

The constraint equations in the inner region are the same with those in the outer region. Plugging $C^{(0)}_{\nu}=0$ into Eq(\ref{inner constraint 1}), we obtain $E^{(0)}_{\nu}=0$. By using $C^{(1)}_{\nu}=\bar{K}^{(1)}_{\nu}=0$, Eq(\ref{inner constraint 2}) becomes 
\begin{equation}
\nu^2 E^{(1)}_{\nu}=-\frac{1}{4i r_{0}}\int^{\infty}_{-\infty} d\omega \omega(\nu-\omega)\phi^{(0)}_{\omega}\phi^{(0)}_{\nu-\omega},
\end{equation}
They are the same with the momentum space expression of Eq(\ref{so-called constraint equations}) and Eq(\ref{so-called constraint equations 2}).
\section*{Acknowledgements}
We would like to thank Sumit R. Das, Alfred D. Shapere, Archisman Ghosh, Lunin Oleg, Willie Merrell, Shiraz Minwalla and Hong Liu for the discussions. This work was partially supported by a National Science Foundation grant NSF-PHY-0855614.

\section*{Appendix}
\appendix
\section{Leading Corrections of the Toy-Model}
\label{Leading Corrections of the Model}
\subsection{Equations of motions}
With Eq(\ref{zeroth_order_metric_and_form_filed_equation}), the leading order Einstein equations for $h(r,v)$, $a(r,v)$ and $k(r,v)$ become
\begin{eqnarray}
\label{Wrr_equation_in_outer_region}
W_{rr}&=&-\frac{2h^{\prime}(r,v)}{r}-\frac{1}{2}(\partial_{v}\phi_{0})^2 \left( \frac{r_{0}^{2}-r^2}{r^2U_{0}(r)} \right)^2=0 \\
\label{Wrv_equation_in_outer_region}
W_{rv}&=&\frac{1}{2r^4} \left(  -12r^4h(r,v) - 2r(r^4+r \epsilon_{0} - \rho^{2}_{0})h^{\prime}(r,v)  \right. \\ \nonumber
&-& \left. 2rk^{\prime}(r,v) +4r^2 \rho_{0} a^{\prime}(r,v) +2k(r,v) + r^2 k^{\prime \prime}(r,v)  \right)  \\ \nonumber
&-&\frac{1}{2}(\partial_{v}\phi_{0})^2 \left( \frac{r_{0}^{2}-r^2}{r^2U_{0}(r)} \right)=0 \\
\label{Wvv_equation_in_outer_region}
W_{vv}&=&-\frac{U_{0}(r)}{2r^4} \left( -12r^4h(r,v) - 2r(r^4+r \epsilon_{0} - \rho^{2}_{0})h^{\prime}(r,v)  \right. \\ \nonumber
&-& \left. 2rk^{\prime}(r,v) +4r^2 \rho_{0} a^{\prime}(r,v) +2k(r,v) + r^2 k^{\prime \prime}(r,v) \right)  \\ \nonumber
&-& \frac{1}{2}(\partial_{v}\phi_{0})^2 + \frac{1}{r^{3}}(2r r^{3}_{0}\dot{E}(v)-2 \rho_{0}r^{2}_{0} \dot{C}(v))=0 \\
\label{Wii_equation_in_outer_region}
W_{ii} &=& -\frac{1}{r^2} ( 6r^4h(r,v) +k(r,v) +r^3 U_{0}(r)h^{\prime}(r,v)\\ \nonumber
&-&r k^{\prime}(r,v) + 2r^2 \rho_{0} a^{\prime}(r,v)) )=0
\end{eqnarray}
,where $W_{ii} \equiv W_{xx}=W_{yy}$ and dots and primes indicate derivatives with respect to $v$ and $r$ respectively. The gauge field equations are
\begin{eqnarray}
\label{scalar_Yv_equation}
Y^{v}&=&-\frac{1}{r^2} \left( \rho_{0} h^{\prime}(r,v) + 2r a^{\prime}(r,v) + r^2 a^{\prime \prime}(r,v) \right)=0, \\
\label{scalar_Yr_equation}
Y^{r}&=&\frac{r^{2}_{0}}{r^2}\dot{C}(v)=0.
\end{eqnarray}
The other components of the Einstein equations and gauge field equations are zero. These are the leading order equations in the naive derivative expansion. This means that $v$-derivatives on $h(r,v)$, $a(r,v)$ and $k(r,v)$ are ignored.

Eq(\ref{scalar_Yr_equation}) shows that there is no dynamics for the charge density. By the initial conditions mentioned in Sec.\ref{Charged Black Brane with Real Massless Scalar Field}, $C(v)=0$. A particular combination of Einstein equations, $W_{rv}U_{0}(r,v)+W_{vv}=0$, gives
\begin{equation}
\label{Energy_density_constraint_equation_of_outer_region}
\dot{E}(v)= \frac{1}{4r_{0}} (\partial_{v}\phi_{(0)}(v))^2 
\end{equation}
This equation indicates that $E(v) \sim O(\varepsilon)$. This justifies that  $E(v)$ in the metric factor $U(r,v)$ is suppressed by $\varepsilon$ to produce the second order terms in Eq(\ref{outer_Scalar_First_order_Equation}).
We solve Eq(\ref{Wrr_equation_in_outer_region}), Eq(\ref{scalar_Yv_equation}), Eq(\ref{Wii_equation_in_outer_region}) to get Eq(\ref{1st_order_scalar_solution_in_outer_region_and_its_backreaction_to_h}), Eq(\ref{1st_order_scalar_solution_in_outer_region_and_its_backreaction_to_a}),Eq(\ref{1st_order_scalar_solution_in_outer_region_and_its_backreaction_to_k}) respectively.
The other Einstein equations are satisfied with the solutions.

\section{Outer Solution in Extremal Limit}
\label{Outer Solution in Extremal Limit}
\subsection{Dilaton solution}
We start with Eq(\ref{1st_order_scalar_solution_in_outer_region}). In the case of extremality, Eq(\ref{1st_order_scalar_solution_in_outer_region}) can have a form of 
\begin{eqnarray}
\label{Divegence_dilaton_1st_order_solution_in_outer_region}
\phi(u,v)&=&\phi_{(0)}(v)+\frac{\partial_{v} \phi_{(0)}(v)}{6(u-1)r_{0}} (1 - \Lambda_{1}(v)) - \frac{\partial_{v}  \phi_{(0)}(v)}{9 r_{0}}ln(u-1) \left( 2 + \Lambda_{1}(v) \right) \\ \nonumber
&+& \frac{\partial_{v}  \phi_{(0)}(v)}{36 r_{0}} \left( (\Lambda_{1}(v)-7)\sqrt{2}tan^{-1}(\frac{u+1}{\sqrt{2}}) + 2(\Lambda_{1}(v)+2)ln(u^2 +2u +3)  \right) +   \Lambda_{2}(v).
\end{eqnarray}
As $u \rightarrow \infty$, a boundary condition that we demand for the dialton is $\left. \phi(u,v) \right|^{u=\infty} = \phi_{(0)}(v)$. This boundary condition yields $\Lambda_{2}(v)=-\frac{i \pi \phi_{0}}{36\sqrt{2}r_{0}}(\Lambda_{1}(v)-7)$.
With this, $\phi(r,v)$ has an asymptotic behavior of
\begin{eqnarray}
\label{Near_Black_brane_HiriZon_expansion_of_the _scalar field}
\phi(u,v)&=&\phi_{(0)}(v)+\frac{\partial_{v}  \phi_{(0)}(v)}{6(u-1)r_{0}} (1 - \Lambda_{1}(v)) - \frac{\partial_{v}  \phi_{(0)}(v)}{9 r_{0}}ln(u-1) \left( 2 + \Lambda_{1}(v) \right) \\ \nonumber
&+& \frac{\partial_{v}  \phi_{(0)}(v)}{36 r_{0}} \left( (\Lambda_{1}(v)-7)\sqrt{2}tan^{-1}(\sqrt{2}) + 2(\Lambda_{1}(v)+2)ln(6) + \frac{\pi}{\sqrt{2}}(7- \Lambda_{1}(v)) \right) \\ \nonumber
&+& O(u-1),
\end{eqnarray}
as $u \rightarrow 1$, near the black brane horizon.

\subsection{Metric and Gauge Field Solution}
As we discussed in Sec.\ref{Matching}, the regularity condition of the dilaton field forces $\Lambda_{1}(v)=1$. In this subsection, we follow this.
For the extremal limit, we set $\epsilon_{0}=2r^{3}_{0}$ and $\rho_{0}=\sqrt{3}r^{2}_{0}$.  Eq(\ref{1st_order_scalar_solution_in_outer_region_and_its_backreaction_to_h}) becomes
\begin{eqnarray}
h(u,v)&=&\bar{h}_{1}(v)-\frac{1}{864}\frac{(\partial_{v}\phi_{(0)}(v))^2}{r^{2}_{0}} \left( \frac{6(u^2-10u-15)}{(u-1)(u^2+2u+3)} - 13\sqrt{2}tan^{-1}(\frac{1+u}{\sqrt{2}}) \right. \\ \nonumber
 &+& \left. 16ln(u-1)-8ln(u^2+2u+3) \right).
\end{eqnarray}
As $u \rightarrow 1$, this can be expanded as 
\begin{eqnarray}
h(u,v)&=& \bar{h}_{1}(v) -\frac{1}{4}\frac{(\partial_{v}\phi_{(0)}(v))^2}{r^{2}_{0}} \left( -\frac{1}{9(u-1)} + \frac{2}{27}ln(u-1) \right. \\ \nonumber
&+&  \left. \frac{1}{216} (8-13\sqrt{2} tan^{-1}(\sqrt{2}) -8ln(6)) +O(u-1)\right).
\end{eqnarray}
The near horizon expansions of Eq(\ref{1st_order_scalar_solution_in_outer_region_and_its_backreaction_to_a}) and Eq(\ref{1st_order_scalar_solution_in_outer_region_and_its_backreaction_to_k}) are also given by
\begin{eqnarray}
a(u,v)&=&\frac{\sqrt{3}}{864}\frac{(\partial_{v}\phi_{(0)}(v))^2}{r_{0}} \left( -8(\frac{2}{u}+1)ln(u-1)  -\frac{30}{u} + \sqrt{2} (\frac{13}{u}-7)tan^{-1}(\frac{1+u}{\sqrt{2}}) \right.\\ \nonumber
&+& \left.4(\frac{2}{u}+1)ln(u^2+2u+3) \right) + \bar{a}_{2}(v)-\frac{\bar{a}_{1}(v)-\sqrt{3}r^{2}_{0}\bar{h}_{1}(v)}{r_{0}u}, \\
k(u,v)&=&-\frac{1}{4}{r^{2}_{0}} (\partial_{v}\phi_{(0)}(v))^{2} \left( \frac{4}{27}(u-1)^2 (u^2+2u+3)ln(u-1) -u^2   \right. \\ \nonumber
&-&  \sqrt{2}tan^{-1}(\frac{u+1}{\sqrt{2}}) \left( -u+\frac{13}{108}(u-1)^2 (u^2+2u+3) \right)  \\ \nonumber
&-& \left. \frac{2(u-1)^2 (u^2 +2u +3)}{27} ln(u^2 +2u +3)  +\frac{1}{18} (u^2 -10u -15)(u-1) \right) \\ \nonumber
&+& r_{0}u\bar{k}_{1}(v) -2 \sqrt{3}r^{2}_{0}\bar{a}_{1}(v) +2r^{4}_{0}(u-1)^{2}(u^2 + 2u +3)\bar{h}_{1}(v),
\end{eqnarray}
The integration constants,$\bar{h}_{1}(v)$, $\bar{a}_{1}(v)$, $\bar{a}_{2}(v)$ and $\bar{k}_{1}(v)$ are determined by the boundary condition (\ref{Outer_region_Boundary_condiTion}). As $u \rightarrow \infty$, the asymptotic expansion of $k(u,v)$ can have terms of $O(u^4)$. These are non-normalizable modes which give deformation of the boundary metric. The terms are removed by imposing 
$\bar{h}_{1}(v)=-\frac{13\sqrt{2}\pi}{1728 r^{2}_{0}}(\partial_{v} \phi_{(0)}(v))^2$. Near $AdS_4$ boundary, $a(u,v)$ presents $O(1)$ and $O(\frac{1}{u})$ terms. The former corrects the chemical potential and the later does the charge density. To eliminate these terms, $\bar{a}_{1}(v)$ and $\bar{a}_{2}(v)$ should be properly chosen as $\bar{a}_{1}(v)=0$ and $\bar{a}_{2}(v)=\frac{7\sqrt{6}\pi}{1728 r_0}(\partial_{v} \phi_{(0)}(v))^2$.

Near the black brane horizon, the behavior of the leading back reactions are given by
\begin{eqnarray}
h(u,v)&=&  -\frac{1}{4}\frac{(\partial_{v}\phi_{(0)}(v))^2}{r^{2}_{0}} \left( -\frac{1}{9(u-1)} + \frac{2}{27}ln(u-1) \right. \\ \nonumber
&+&  \left. \frac{1}{216} (8+\frac{13\sqrt{2}\pi}{2}-13\sqrt{2} tan^{-1}(\sqrt{2}) -8ln(6)) +O(u-1)\right), \\ 
a(u,v)&=&\frac{\sqrt{3}}{864}\frac{(\partial_{v}\phi_{(0)}(v))^2}{r_{0}} \left( -8ln(u-1) (3-2(u-1)+O(u-1)^2 ) -30  \right. \\ \nonumber
&+& \left. 6\sqrt{2}tan^{-1} (\sqrt{2}) - 3\sqrt{2}\pi + 12 ln(6) \right.\\ \nonumber
&+&\left.\left( 40-13\sqrt{2}tan^{-1}(\sqrt{2})-8ln(6)+\frac{13\sqrt{2}\pi}{2} \right)(u-1) + O(u-1)^{2}\right),\\
k(u,v)&=& -\frac{1}{4}{r^{2}_{0}} (\partial_{v}\phi_{(0)}(v))^{2} \left( \frac{8}{9} (u-1)^2 ln(u-1) -1 + \sqrt{2} tan^{-1} (\sqrt{2}) \right. \\ \nonumber
&+& \left. 
  ( \sqrt{2}tan^{-1}(\sqrt{2})-3)(u-1) + \left( \frac{13\sqrt{2}\pi}{36}-\frac{11}{9}-\frac{4ln(6)}{9} \right.\right. \\ \nonumber
&-&\left.\left. \frac{13tan^{-1}(\sqrt{2})}{9\sqrt{2}} \right)(u-1)^{2}\right)+r_{0} \bar{k}_{1}(v) +r_{0} \bar{k}_{1}(v) (u-1) + O(u-1)^{3}.
\end{eqnarray}

\section{Equations in Extremal Backgrounds with $v$-derivative Retained}
\label{Master Equations in Extremal Backgroud}
In this section, we develop the Einstein equations(\ref{Wequation_in_scalar_field}) and the gauge field equations(\ref{Yequation_in_scalar_field}) without ignoring $v$-derivatives. We start with Eq(\ref{zeroth_order_metric_and_form_filed_equation}). 
The only assumption in this section is that the equations are linear in $h(u,v)$, $a(u,v)$ and $k(u,v)$. The Einstein equations are
\begin{eqnarray}
\label{Master Equations in Extremal Background_Wrr}
r^{2}_{0}W_{rr}&=&-\frac{2h^{\prime}(u,v)}{u}-\frac{1}{2}\partial_{u}\phi(u,v)\partial_{u}\phi(u,v)=0, \\
W_{rv}&=&\frac{1}{2u^4} \left(  -12u^4h(u,v) + 2u(3-u^4-2u)h^{\prime}(u,v)+ \frac{2u^{4}}{r_{0}}\dot{h}^{\prime}(u,v) \right. \\ \nonumber
&+&\frac{4\sqrt{3}u^2}{r_{0}}  a^{\prime}(u,v) + \frac{1}{r^{4}_{0}}\left(2k(u,v)- 2uk^{\prime}(u,v) + u^2 k^{\prime \prime}(u,v)  \right),  \\ \nonumber
&-&\frac{1}{2r_{0}}\partial_{v}\phi(u,v)\partial_{u}\phi(u,v)=0, \\
\label{Master Equations in Extremal Background_W-bar}
\bar{W} &\equiv& W_{vv}+\left(\frac{r^{4}-4rr^{3}_{0}+3r^{4}_{0}}{r^2}\right)W_{rv} \\ \nonumber
&=&\frac{r_{0}}{u^3}\left(-2\sqrt{3}\dot{C}(v)+2u\dot{E}(v)-2(u^{4}-4u+3)\dot{h}(u,v)+\frac{\dot{k}(u,v)}{r^{4}_{0}}\right) \\ \nonumber
&-&\frac{1}{2}\partial_{v}\phi(u,v)\partial_{v}\phi(u,v)-r_{0}\left(\frac{u^{4}-4u+3}{2u^2}\right)\partial_{u}\phi(u,v)\partial_{v}\phi(u,v)=0, \\
\frac{r^{2}}{r^{2}_{0}}W_{ii}&=&-6u^{4}h(u,v)-u(u^{4}-4u+3)h^{\prime}(u,v)-\frac{2\sqrt{3}u^{2}}{r_{0}}a^{\prime}(u,v) \\ \nonumber
&+&\frac{1}{r^{4}_{0}}\left(uk^{\prime}(u,v)-k(u,v)\right)=0,
\end{eqnarray}
where $u=\frac{r}{r_{0}}$, the rescaled radial coordinate again. The primes and dots denote derivatives with respect to $u$ and $v$ respectively. Gauge field equations are
\begin{eqnarray}
\frac{r^{2}}{r^{2}_{0}}Y^{r}&=&\dot{C}(v)+\sqrt{3}\dot{h}(u,v)+\frac{u^{2}}{r_{0}}\dot{a}^{\prime}(u,v)=0, \\
r^{2}Y^{v}&=&-2ua^{\prime}(u,v)-\sqrt{3}r_{0}h^{\prime}(u,v)-u^{2}a^{\prime\prime}(u,v)=0.
\end{eqnarray}

As discussed in the beginning of Sec.\ref{Divergence Resolution of Massless Scalar Field}, for introducing the scaling(\ref{scaling_limit}) the equations in position space of $v$ should be transformed to the momentum space. It is worth showing how one gets an expression Eq(\ref{Master Equations in Extremal Background_Wrr}) in momentum space, for example. We define the Fourier transform of $h(u,v)$ by
\begin{equation}
\label{define Fourier transformations_h}
h(u,v)=\int^{\infty}_{-\infty} h_{\omega}(u)e^{i \omega v} d\omega.
\end{equation} 
Substituting of Eq(\ref{define Fourier transformations_scalar}) and Eq(\ref{define Fourier transformations_h}) into Eq(\ref{Master Equations in Extremal Background_Wrr}) and acting an integral operator $\frac{1}{2\pi} \int^{\infty}_{-\infty} e^{-i \nu v} dv$ on it, we get 
\begin{equation}
\label{Einstein equations in momentum space_Wrr_in_appendix}
r^{2}_{0}W_{rr}=-\frac{2h_{\nu}^{\prime}(u)}{u}-\frac{1}{2}\int^{\infty}_{-\infty} d \omega \partial_{u}\phi_{\nu-\omega}(u)\partial_{u}\phi_{\omega}(u)=0.
\end{equation}
To deal with other equations, we define Fourier transforms of $a(u,v)$ and $k(u,v)$ as that of $h(u,v)$. For $E(v)$ and $C(v)$, 
\begin{eqnarray}
\label{Fourier_transformation_of_C_and_E}
E(v)=\int^{\infty}_{-\infty} e^{i\omega v} E_{\omega} d \omega, \\ \nonumber
C(v)=\int^{\infty}_{-\infty} e^{i\omega v} C_{\omega} d \omega.
\end{eqnarray}
With these, the other Einstein equations in momentum space are given by
\begin{eqnarray}
\label{Einstein equations in momentum space_Wrv_in_appendix}
W_{rv}&=&\frac{1}{2u^4} \left(  -12u^4h_{\nu}(u) + 2u(3-u^4-2u)h_{\nu}^{\prime}(u)+ i\nu\frac{2u^{4}}{r_{0}}h_{\nu}^{\prime}(u) \right. \\ \nonumber
&+&\frac{4\sqrt{3}u^2}{r_{0}}  a_{\nu}^{\prime}(u) + \frac{1}{r^{4}_{0}}\left( 2k_{\nu}(u)- 2uk_{\nu}^{\prime}(u) + u^2 k_{\nu}^{\prime \prime}(u)\right)  \\ \nonumber
&-&\frac{1}{2r_{0}}\int^{\infty}_{-\infty} d \omega i(\nu-\omega)\phi_{\nu-\omega}(u)\partial_{u}\phi_{\omega}(u)=0, \\
\label{Einstein equations in momentum space_barW_in_appendix}
\bar{W} &\equiv& W_{vv}+\left(\frac{r^{4}-4rr^{3}_{0}+3r^{4}_{0}}{r^2}\right)W_{rv} \\ \nonumber
&=&\frac{i\nu r_{0}}{u^3}\left(-2\sqrt{3}C_{\nu}+2uE_{\nu}-2(u^{4}-4u+3)h_{\nu}(u)+\frac{k_{\nu}(u)}{r^{4}_{0}}\right) \\ \nonumber
&-&r_{0}\left(\frac{u^{4}-4u+3}{2u^2}\right)\int^{\infty}_{-\infty} d \omega i(\nu-\omega)\phi_{\nu-\omega}(u)\partial_{u}\phi_{\omega}(u) \\ \nonumber
&+&\frac{1}{2}\int^{\infty}_{-\infty} d \omega (\nu-\omega)\omega \phi_{\nu-\omega}(u)\phi_{\omega}(u)=0, \\
\label{Einstein equations in momentum space_Wii_in_appendix}
\frac{r^{2}}{r^{2}_{0}}W_{ii}&=&-6u^{4}h_{\nu}(u)-u(u^{4}-4u+3)h_{\nu}^{\prime}(u)-\frac{2\sqrt{3}u^{2}}{r_{0}}a_{\nu}^{\prime}(u) \\ \nonumber
&+&\frac{1}{r^{4}_{0}}\left(uk^{\prime}_{\nu}(u)-k_{\nu}(u)\right)=0.
\end{eqnarray}
The gauge field equations become
\begin{eqnarray}
\label{gauge field equations in momentum space_Yr_in_appendix}
\frac{r^{2}}{r^{2}_{0}}Y^{r}&=&i\nu\left(C_{\nu}+\sqrt{3}h_{\nu}(u)+\frac{u^{2}}{r_{0}}a_{\nu}^{\prime}(u)\right)=0, \\
\label{gauge field equations in momentum space_Yv_in_appendix}
r^{2}Y^{v}&=&-2ua_{\nu}^{\prime}(u)-\sqrt{3}r_{0}h_{\nu}^{\prime}(u)-u^{2}a_{\nu}^{\prime\prime}(u)=0.
\end{eqnarray}


\section{Counting Power of $\varepsilon$}
\label{Counting Order in Perturbation Theory}
In this section, we argue the parametric order of the inner region solutions in $\varepsilon$. The basic idea is that we transfrom the inner region solutions to the position space and check their powers of $\varepsilon$. For a simple example, we discuss the dilaton solution.
We design the zeroth order dilaton solution in position space as in Eq(\ref{parametric_form_of_zeroth_dilaton_field}). By Fourier transformation, we obtain its expression in frequency space as
\begin{equation}
\phi^{(0)}_{\omega}=\frac{1}{2\pi}\int^{\infty}_{-\infty} e^{-i \omega v}f(\frac{\varepsilon v}{r_{0}}) dv.
\end{equation}
Scaling of the integration variable $v$ as $\tau=\varepsilon v$ takes this expression to 
\begin{equation}
\label{a formal expression of phi with f}
\phi^{(0)}_{\omega}=\frac{1}{2\pi \varepsilon}\int^{\infty}_{-\infty} e^{-i  \frac{\omega}{\varepsilon}\tau}f(\frac{\tau}{r_{0}}) d\tau \equiv \frac{1}{\varepsilon} g(\frac{\omega}{\varepsilon}),
\end{equation}
where $g(\frac{\omega}{\varepsilon})$ becomes an $O(1)$ function. By observing Eq(\ref{a formal expression of phi with f}) and dilaton solution in the inner region(\ref{the_final_result_of_dialton}), one can recognize that each subleading correction to the dialton field in momentum space can be written as
\begin{equation}
\phi^{(i)}_{(in)\omega} \equiv \frac{1}{\varepsilon} g(\frac{\omega}{\varepsilon}) \left(h_{(i)}(\xi) +a_{(i)}(\xi)ln(\nu) \right),
\end{equation}
where $\phi^{(i)}_{(in)\omega}$ denotes $i$th order correction in small frequency to the dilaton field. $h_{(i)}(\xi)$ and $a_{(i)}(\xi)$ are functions of $\xi$ only.
It turns out that the terms multiplying $a_{(i)}(\xi)$ produce terms which are proportional to $\varepsilon^{i}ln(\varepsilon)$ in the position space. It is obscure to count power of $\varepsilon$ of these terms. In this discussion, we exclude these. The perturbation expansion of the dilaton solution becomes a form of
\begin{equation}
\label{nu counting in the momentum space}
\phi_{(in)\omega}=\frac{1}{\varepsilon} g(\frac{\omega}{\varepsilon}) \left( 1 + \nu h_{(1)}(\xi) + \nu^{2}  h_{(2)}(\xi)... \right),
\end{equation}
up to the logarithmic terms. Fourier transformation defined in Eq(\ref{define Fourier transformations_scalar}) takes this expression to position space, which is given by
\begin{equation}
\phi(u,v)=\int^{\infty}_{-\infty} e^{i \omega v}  d \omega \frac{1}{\varepsilon} g(\frac{\omega}{\varepsilon}) \left( 1 + \omega h_{(1)}(\xi) + \omega^{2}  h_{(2)}(\xi)... \right).
\end{equation}
Again we rescale the integration variable $\omega$ as $\omega=\varepsilon \bar{\omega}$, then the expression becomes
\begin{eqnarray}
\label{dilaton_expansioN_in_position_space_in_appendix}
\phi(u,v)&=&\int^{\infty}_{-\infty} e^{i \varepsilon \bar{\omega} v}  d \bar{\omega}  g(\bar{\omega}) \left( 1 + \varepsilon \bar{\omega} h_{(1)}(\xi) + \varepsilon^{2} \bar{\omega}^{2}  h_{(2)}(\xi)... \right) \\ \nonumber
&\equiv& F^{(0)}(\varepsilon v) + (-i)\varepsilon F^{(0)\prime}(\varepsilon v) h_{(1)}(\xi) + (-i)^{2}\varepsilon^{2} F^{(0)\prime\prime}(\varepsilon v) h_{(2)}(\xi),
\end{eqnarray}
where $F^{(0)}(\varepsilon v) = \int^{\infty}_{-\infty} e^{i \varepsilon \bar{\omega} v}  d \bar{\omega}  g(\bar{\omega})$ and the prime indicates derivative with respect to its argument. The property of $F^{(0)}(\varepsilon v)$ as noted in Eq (\ref{derivative_argument_parametric_form_of_zeroth_dilaton_field}) shows that $F^{(0)}(\varepsilon v)$ and its derivatives with its argument
are $O(1)$ functions. Compare Eq(\ref{nu counting in the momentum space}) to Eq(\ref{dilaton_expansioN_in_position_space_in_appendix}). This shows that counting power of $\nu$ in the momentum space is the same as the counting power of $\varepsilon$ in the position space.

We apply this argument to the back reactions in the inner region. For the simplest case, let us check $h^{(1)}_{(in)\nu}(\xi)$ with Eq(\ref{The solutions up to the first subleading_of_2nd_order_equation_of_H1}). This contains derivative of the dilaton field, which can be expressed as $\phi^{(1)\prime}_{(in)\omega}(\frac{\omega}{\nu}\xi)=\frac{1}{\varepsilon}f(\frac{\omega}{\varepsilon})k(\frac{\omega}{\nu}\xi)$. 
We switch Eq(\ref{The solutions up to the first subleading_of_2nd_order_equation_of_H1}) to position space with Fourier transformation as defined in Eq(\ref{define Fourier transformations_h}) and rescale integration variables as in the discussion of the dilaton. This time, we scale $\omega$ as well as $\nu$ in Eq(\ref{The solutions up to the first subleading_of_2nd_order_equation_of_H1}). Then, we obtain following expression:
\begin{eqnarray}
\label{varepsilon counting in the position space of h}
h^{(1)}_{(in)}(v,\xi)&=&\varepsilon \int^{\xi}\frac{\xi^{\prime 2}d\xi^{\prime}}{4\nu^3}\int^{\infty}_{-\infty}e^{i \bar{\nu} \varepsilon v}d\bar{\nu}d\bar{\omega} \bar{\omega}^{2}(\bar{\nu}-\bar{\omega})^{2}f(\bar{\omega})f(\bar{\nu}-\bar{\omega})g(\frac{\bar{\omega}}{\bar{\nu}}\xi)g(\frac{\bar{\nu}-\bar{\omega}}{\bar{\nu}}\xi) \\ \nonumber
&\equiv& \varepsilon G(\varepsilon v),
\end{eqnarray}
where $G(\varepsilon v)$ is an $O(1)$ function. Consequently, it turns out that $h^{(1)}_{(in)}(v,\xi)$ is in the first order in $\varepsilon$ in position space. Comparing Eq(\ref{varepsilon counting in the position space of h}) to Eq(\ref{The solutions up to the first subleading_of_2nd_order_equation_of_H1}), one can recognize that the correct power counting of the small frequency in the momentum space is to count not only $\nu$ but also $\omega$ and $\nu-\omega$ in the integrand of Eq(\ref{The solutions up to the first subleading_of_2nd_order_equation_of_H1}). By the the same argument, we get
\begin{eqnarray}
a^{(1)}_{(in)}(v,\xi) &\sim& k^{(1)}_{(in)}(v,\xi) \sim O(\varepsilon), \\ \nonumber
h^{(2)}_{(in)}(v,\xi) &\sim&  a^{(2)}_{(in)}(v,\xi) \sim k^{(2)}_{(in)}(v,\xi) \sim O(\varepsilon^{2}), \\ \nonumber
a^{(3)}_{(in)}(v,\xi) &\sim& k^{(3)}_{(in)}(v,\xi) \sim O(\varepsilon^{3}) {\rm\ \ and\ \ } k^{(4)}_{(in)}(v,\xi) \sim O(\varepsilon^{4}).
\end{eqnarray}


\end{document}